\documentclass[authorversion,nonacm]{acmart}
\AtBeginDocument{%
  \providecommand\BibTeX{{%
    \normalfont B\kern-.05em{\scshape i\kern-.025em b}\kern-.08em\TeX}}}
\usepackage[libertine]{newtxmath}
\usepackage{stfloats}
   \graphicspath{{./pdf/}{./png/}{./jpeg/}{./eps/}{./pst/}}
   \DeclareGraphicsExtensions{.pdf,.png,.jpeg,.eps}
\usepackage{enumitem}
\definecolor{DarkGreen}{rgb}{0,0.3333,0}

\renewcommand{\d}{{\mathrm d}}
\newcommand{\half}{{\frac{1}{2}}}
\newcommand{\sgn}{{\mathrm{sgn}}}
\newcommand{\Seps}{{{\mathcal{S}}_{\scriptscriptstyle \varepsilon}}}
\DeclareMathOperator\erfc{erfc}
\setcopyright{rightsretained}
\copyrightyear{2020}
\acmYear{2020}
\begin{document}
\title[Pileup Effect and Intermittently Nonlinear Filtering in Synthesis of Covert Communications]{Utilizing Pileup Effect and Intermittently Nonlinear Filtering in Synthesis of Covert and Hard-to-Intercept Communication Links}
\author{Alexei V. Nikitin}
\affiliation{%
  \institution{Nonlinear LLC}
  \city{Wamego}
  \state{Kansas}
  \country{USA}}
\email{avn@nonlinearcorp.com}

\author{Ruslan L. Davidchack}
\affiliation{%
  \institution{University of Leicester}
  \city{Leicester}
  \country{UK}}
\email{rld8@leicester.ac.uk}
\renewcommand{\shortauthors}{Nikitin and Davidchack}
\begin{abstract}
We outline an approach to physical-layer steganography where the transmitted low-power stego messages are statistically indistinguishable from the Gaussian component of the channel noise (e.g. the thermal noise) observed in the same spectral band, and thus the channel noise itself serves as an effective cover signal. We also demonstrate how the apparent spectral and temporal properties of transmitted additional, higher-power cover signals (including those using the existing communication protocols) can be made to match those of the low-power stego payload and the Gaussian noise, providing extra layers of obfuscation for both the cover and the stego messages. We further illustrate how a specific combination of linear and nonlinear filtering can be used for effective separation of the cover, payload, and/or ``friendly jamming" signals even when all transmissions have essentially the same spectral characteristics as well as temporal and amplitude structures, and when there are no explicit differences in the spectral and/or temporal allocations for the cover and the stego messages.
\end{abstract}

\begin{teaserfigure}
  \centering
  \includegraphics[width=.666\textwidth]{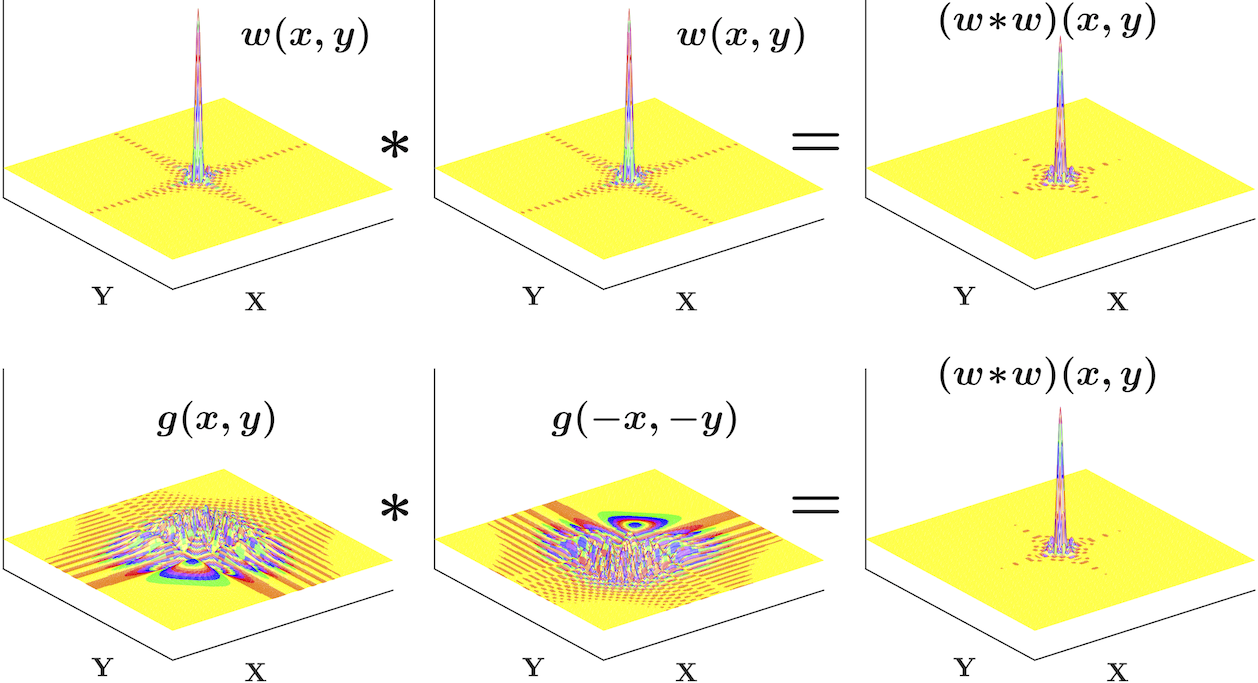}
  \caption{\boldmath Matched filter pairs with identical spectral characteristics and combined responses, but significantly different time and/or spatial supports, can also be constructed for multidimensional filters, for example spatial 2D ($g_i(x,y)$) and/or spatio-temporal 3D ($g_i(x,y,t)$) filters for image and/or video processing.}
  \Description{Matched filter pairs with identical spectral characteristics and combined responses, but significantly different time and/or spatial supports, can also be constructed for multidimensional filters, for example spatial 2D ($g_i(x,y)$) and/or spatio-temporal 3D ($g_i(x,y,t)$) filters for image and/or video processing.}
  \label{fig:teaser}
\end{teaserfigure}
\maketitle

\section{Introduction}
To meet the undetectability requirement, in a steganographic system the stego signals should be statistically indistinguishable from the cover signals. For physical layer transmissions, this can perhaps be enhanced by requiring that the payload and the cover have the same bandwidth and spectral content, the same apparent temporal and amplitude structures, and that there are no explicit differences in the spectral and/or temporal allocations for the cover signals and the payload messages.

For a mixture of such signals, neither linear nor nonlinear filtering alone can separate the signals. Favorably, however, linear filtering can significantly, and differently, affect the temporal and amplitude structure of many natural and the majority of {\itshape technogenic\/} (man-made) signals. For example, such filtering can often convert the amplitude distribution of a pulse train from super-Gaussian into apparently Gaussian and/or sub-Gaussian, and {\itshape vice versa\/}. On the other hand, a nonlinear filter is capable of disproportionately affecting spectral densities of signals with distinct temporal and/or amplitude structures even when the signals have the same spectral content. Therefore, a proper synergistic combination of linear and nonlinear filtering can be employed to effectively separate such ``indistinguishable" cover and stego signals.

\subsection{Channel Noise as Cover Signal} \label{subsec:channel cover}
The very existence of a detectable carrier (cover signal) may be a dead giveaway for the stego payload. For example, a simple presence of a sheet of paper implies the possibility of a message written in invisible ink. Therefore, the best steganography should be ``carrier-less," when the payload is covertly embedded into something ``ever-present." In the physical layer, such ``ideal" and unidentifiable cover signal is the channel noise. Such noise always includes the ever-present thermal noise as one of its components, and may also comprise other (in general, non-Gaussian) natural and/or technogenic (man-made) components. Then, if the stego payload ``pretends" to be Gaussian, and its power is small enough to be well within the natural variations of the channel noise, any physically available band can be used to carry a virtually undetectable covert message.

In this paper, we outline an approach to physical-layer steganography where the transmitted low-power stego messages are statistically indistinguishable from the Gaussian component of the channel noise (e.g. the thermal noise) observed in the same spectral band, and thus the channel noise itself serves as an effective cover signal. We also demonstrate how the apparent spectral and temporal properties of transmitted additional, higher-power cover signals (including those using the existing communication protocols) can be made to match those of the low-power stego payload and the Gaussian noise, providing extra layers of obfuscation for both the cover and the stego messages. We further illustrate how a specific combination of linear and nonlinear filtering can be used for effective separation of the cover, payload, and/or ``friendly jamming" signals even when all transmissions have effectively the same spectral characteristics as well temporal and amplitude structures, and when there are no explicit differences in the spectral and/or temporal allocations for the cover and the stego messages.

Throughout the paper, while keeping the presentation rather abbreviated, we attempt to provide sufficient amount of detail required for subsequent practical development of this approach.

\begin{figure}[!b]
  \centering
  \includegraphics[width=.666\linewidth]{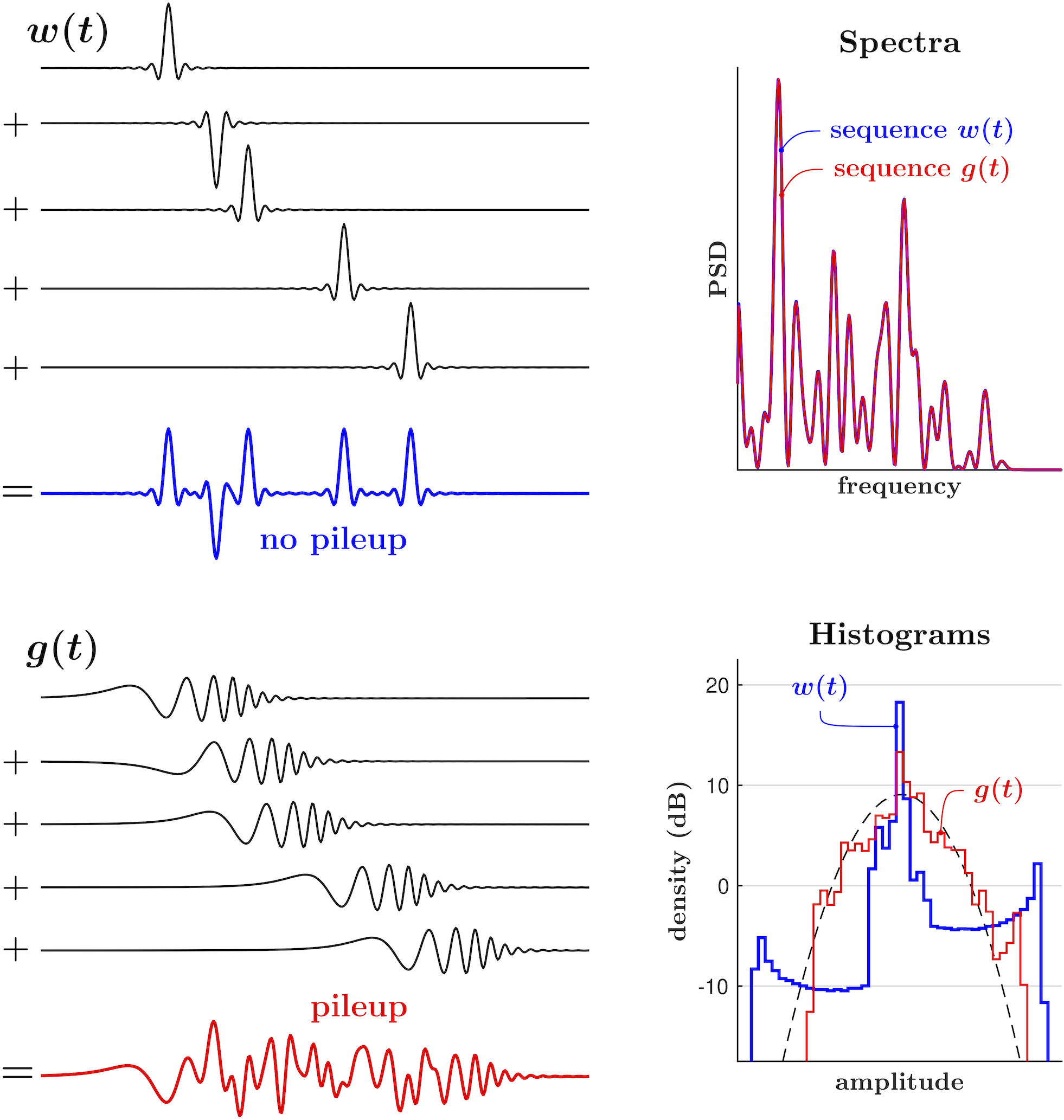}
  \caption{Illustration of pileup effect: When ``width" of pulses becomes greater than distance between them, pulses begin to overlap and interfere with each other. For pulses with same spectral content, PSDs of pulse sequences are identical, yet their temporal and amplitude structures are substantially different.}
  \Description{Illustration of pileup effect: When ``width" of pulses becomes greater than distance between them, pulses begin to overlap and interfere with each other. For pulses with same spectral content, PSDs of pulse sequences are identical, yet their temporal and amplitude structures are substantially different.}
  \label{fig:PPileup}
\end{figure}

\section{Mimicking Function of Pileup Effect}
A pulse train~$p(t)$ is simply a sum of pulses with the same shape (impulse response)~$w(t)$, same or different amplitudes~$a_k$, and distinct arrival times~$t_k$: $p(t) = \sum_k a_k w(t-t_k)$. When the width of the pulses in a train becomes greater than the distance between them, the pulses begin to overlap and interfere with each other. This is illustrated in Fig.~\ref{fig:PPileup}: For the same arrival times, the pulses in the sequence consisting of the narrow pulses~$w(t)$ remain separate, while the wider (more ``spread out") pulses~$g(t)$ are ``piling up on top of each other." In this example, $w(t)$ and $g(t)$ have the same spectral content, and thus the power spectral densities (PSDs) of the pulse sequences are identical. However, the ``pileup effect" causes the temporal and amplitude structures of these sequences to be substantially different. For a random pulse train, when the ratio of the bandwidth and the pulse arrival rate becomes significantly smaller than the time-bandwidth product (TBP) of a pulse, the pileup effect causes the resulting signal to become effectively Gaussian~\cite[e.g.]{Nikitin98ppileup}, making it impossible to distinguish between the individual pulses.

Indeed, let~$\hat{p}(t)$ be an ``ideal" pulse train: $\hat{p}(t) = \sum_k a_k\delta(t-t_k)$, where $\delta(x)$ is the Dirac $\delta$-function~\cite{Dirac58principles}. The {\itshape moving average\/} of this ideal train in a boxcar window of width~$2T$ can be represented by the convolution integral
\begin{equation} \label{eq:moving average}
  \overline{p}(t) = \int_{-\infty}^\infty \!\!\d{s}\, \frac{\theta(t\!+\!T)-\theta(t\!-\!T)}{2T}\, \hat{p}(t\!-\!s)\,,
\end{equation}
where $\theta(x)$ is the Heaviside unit step function~\cite{Bracewell2000Fourier}. At any given time~$t_i$, the value of $\overline{p}(t_i)$ is proportional to the sum of~$a_k$ for the pulses that occur within the interval~$[t_i\!-\!T,t_i\!+\!T]$. Then, if the amplitudes $a_k$ and/or the interarrival times~$t_{k+1}-t_k$ are independent and identically distributed (i.i.d.) random variables with finite mean and variance, it follows from the central limit theorem (CLT)~\cite[e.g.]{Aleksandrov56mathematics} that the distribution of $\overline{p}(t_i)$ approaches Gaussian for a sufficiently large interval~$[-T,T]$.

If we replace the boxcar weighting function in~(\ref{eq:moving average}) with an arbitrary moving window~$w(t)$, then~(\ref{eq:moving average}) becomes a {\itshape weighted\/} moving average
\begin{equation} \label{eq:weighted moving average}
  p(t) = \int_{-\infty}^\infty \!\!\d{s}\, w(t)\, \hat{p}(t\!-\!s) = (\hat{p}\!\ast\!w)(t) = \sum_k a_k w(t-t_k)\,,
\end{equation}
which is a ``real" pulse train with the impulse response~$w(t)$.\footnote{In~(\ref{eq:weighted moving average}) and throughout the paper the asterisk denotes convolution.} If~$w(t)$ is normalized so that~$\int_{-\infty}^\infty \d{s}\, w(s) = 1$, $w(t)$ is an {\itshape averaging\/} (i.e. lowpass) filter. Then, if~$w(t)$ has both the bandwidth and the time-bandwidth product (TBP) similar to that of the boxcar pulse of width~$2T$, the distribution of $p(t_i)$ would be similar to that of~$\overline{p}(t_i)$ (e.g. Gaussian for a sufficiently large~$T$).

\begin{figure}[!b]
  \centering
  \includegraphics[width=.666\linewidth]{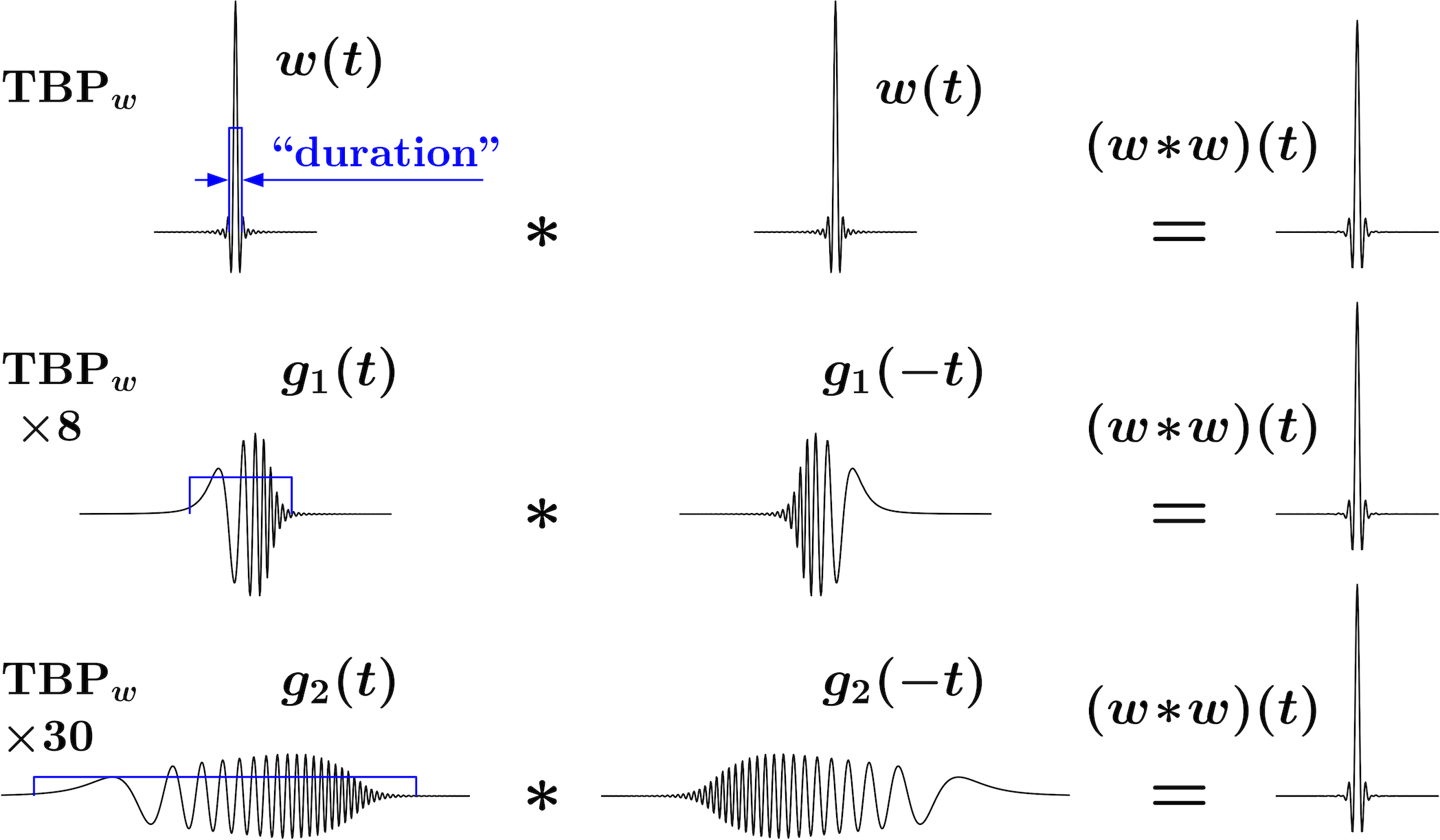}
  \caption{Pairs of matched filters with different time-bandwidth products, but same frequency responses and same ``combined" impulse response. In this example, $w(t)$ is root-raised-cosine filter, and thus $(w\!\ast\! w)(t)$ is raised-cosine filter.}
  \Description{Pairs of matched filters with different time-bandwidth products, but same frequency responses and same ``combined" impulse response. In this example, $w(t)$ is root-raised-cosine filter, and thus $(w\!\ast\! w)(t)$ is raised-cosine filter.}
  \label{fig:TBPs}
\end{figure}

\subsection{TBP of Filter in Context of Pileup Effect} \label{subsec:TBP in context}
There are various ways to define the ``time duration" and the ``bandwidth" of a pulse. This can lead to a significant ambiguity in the definitions of the TBPs, especially for filters with complicated temporal structures and/or frequency responses. However, in the context of a mimicking function of the pileup effect, our main concern is the change in the TBP that occurs only due to the change in the temporal structure of a filter, without the respective change in its spectral content. For a single pulse~$w(t)$, its peak-to-average power ratio (PAPR) can be expressed as
\begin{equation} \label{eq:PAPR}
  {\rm PAPR}_w = \frac{\max\left(w^2(t)\right)}{\frac{1}{T_2-T_1} \int_{T_1}^{T_2} \d{t}\, w^2(t)}\,,
\end{equation}
where the interval~$[T_1,T_2]$ includes the effective time support of~$w(t)$. Then for filters with the same spectral content and the impulse responses~$w(t)$ and~$g(t)$, the ratio of their TBPs can be expressed as the reciprocal of the ratio of their PAPRs,
\begin{equation} \label{eq:TBP}
  \frac{{\rm TBP}_g}{{\rm TBP}_w} = \frac{\max\left(w^2(t)\right)}{\max\left(g^2(t)\right)} = \frac{{\rm PAPR}_w}{{\rm PAPR}_g}\,,
\end{equation}
where the PAPRs are calculated over a sufficiently long time interval that includes the effective time support of both filters.

Note that from~(\ref{eq:TBP}) it follows that, among all possible pulses with the same spectral content, the one with the smallest TBP will contain a dominating large-magnitude peak. Hence any reasonable definition of a finite TBP for a particular filter with a given frequency response allows us to obtain comparable numerical values for the TBPs of all other filters with the same frequency response, regardless of their temporal structures. For example, defining the ``time duration" of the pulses~$g_1(t)$ and~$g_2(t)$ shown in Fig.~\ref{fig:TBPs} can be challenging. On the other hand, a ``reasonable" definition of the duration of the root-raised-cosine pulse~$w(t)$ can be given as~$2T_{\rm s}$, where~$T_{\rm s}$ is the the reciprocal of the symbol-rate parameter of the pulse. Then defining the bandwidth by the 3\,dB corner frequency (i.e. $\Delta{B}=(2T_{\rm s})^{-1}$) leads to ${\rm TBP}_w=1$. 

\begin{figure}[!b]
  \centering
  \includegraphics[width=.666\linewidth]{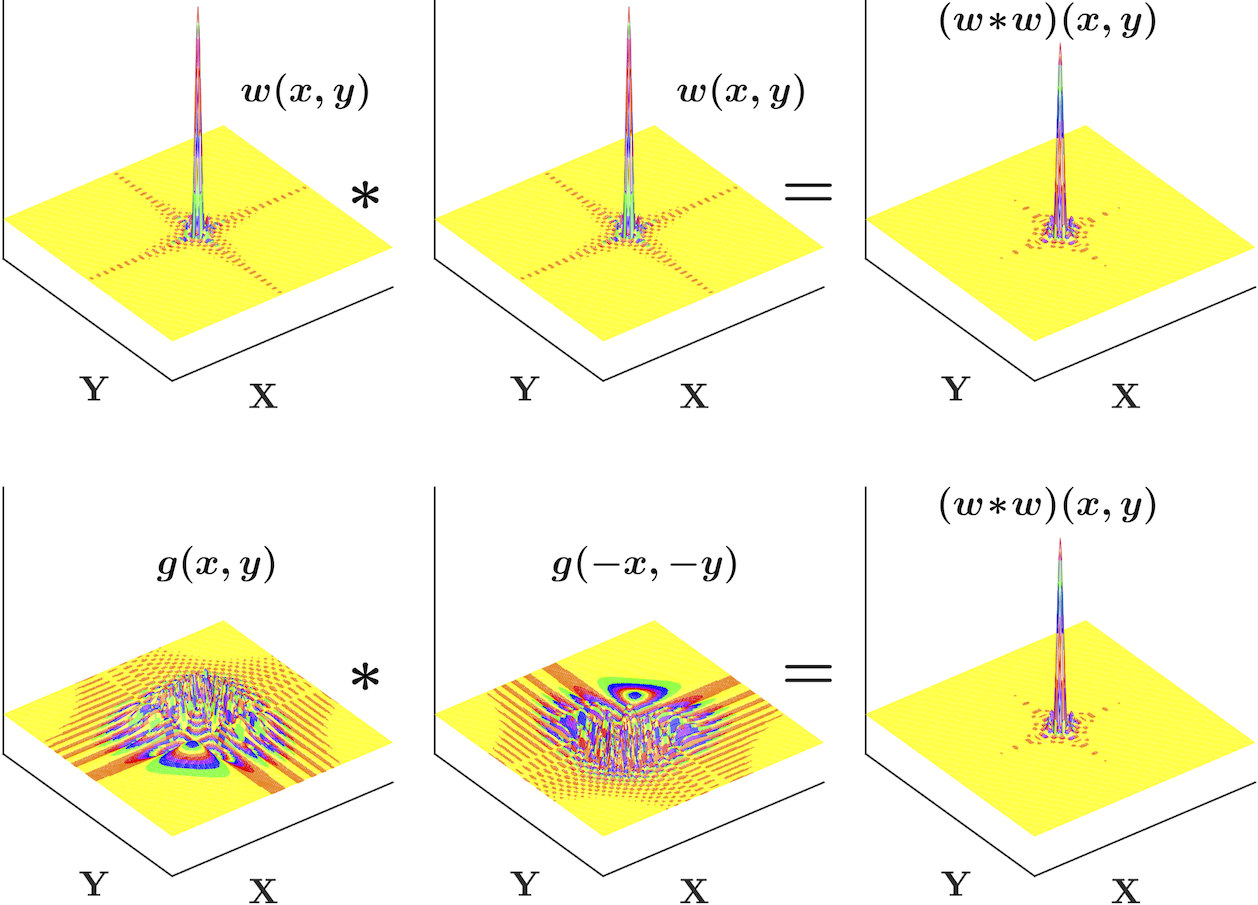}
  \caption{Example of Fig.~\ref{fig:TBPs} extended to two dimensions.}
  \Description{Example of Fig.~\ref{fig:TBPs} extended to two dimensions.}
  \label{fig:TBPs 2D}
\end{figure}
\begin{figure*}[!t]
  \centering
  \includegraphics[width=\textwidth]{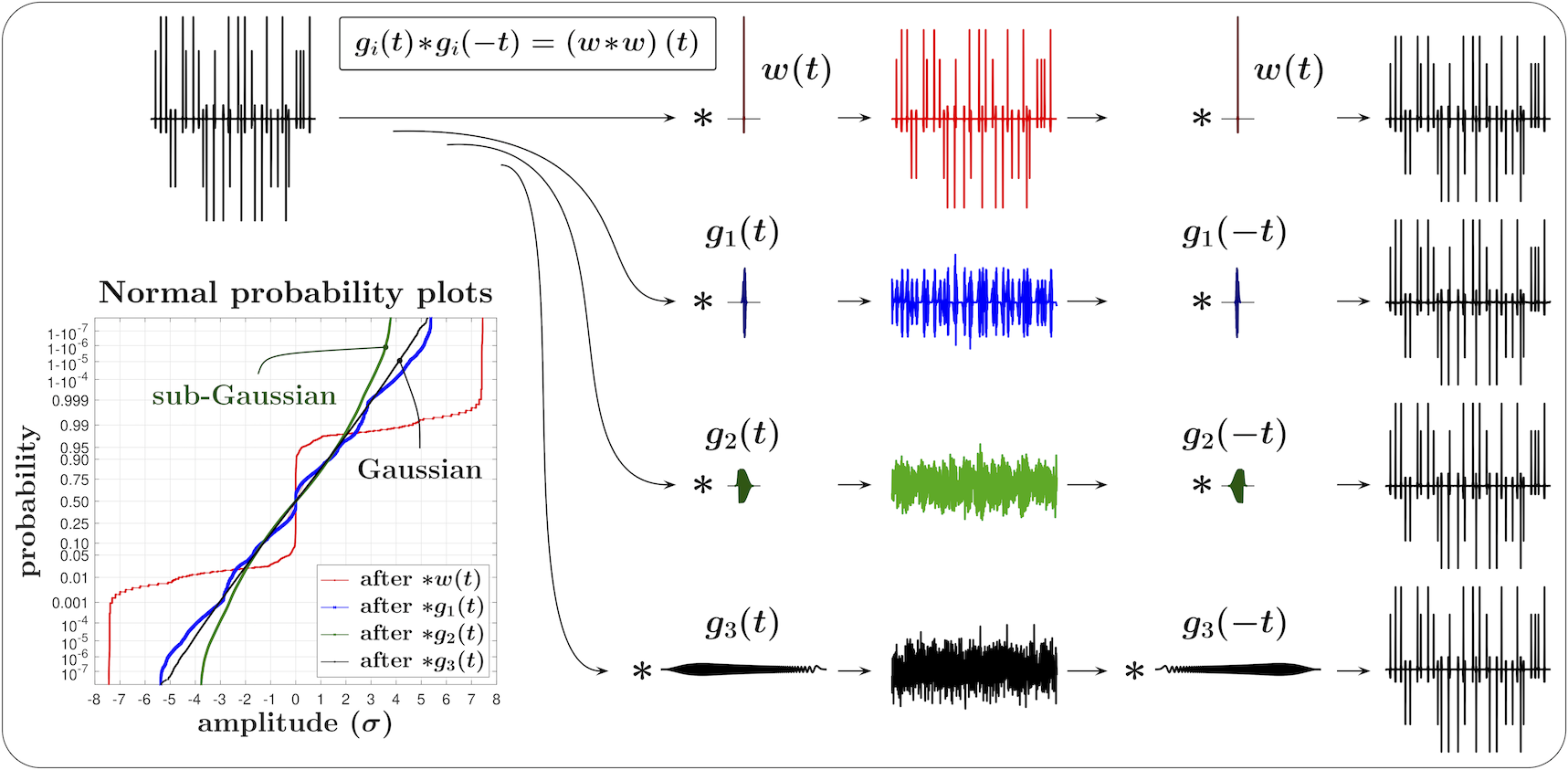}
  \caption{Using pileup effect for obfuscation of temporal and amplitude structure of transmitted signal. In transmitter, filtering with large-TBP filter reduces crest factor of pulse train, making it appear as Gaussian or sub-Gaussian. In receiver, filtering with matched filter restores signal's distinct temporal and amplitude structure.}
  \Description{Using pileup effect for obfuscation of temporal and amplitude structure of transmitted signal. In transmitter, filtering with large-TBP filter reduces crest factor of pulse train, making it appear as Gaussian or sub-Gaussian. In receiver, filtering with matched filter restores signal's distinct temporal and amplitude structure.}
  \label{fig:ppileup}
\end{figure*}

Given a ``seed" pulse~$w(t)$, perhaps the easiest way to construct a pulse~$g(t)$ with the same spectral content but a different TBP is to filter~$w(t)$ with an all-pass filter, for example, a linear or nonlinear chirp with a flat frequency response. Then the convolutions of $w(t)$ and $g(t)$ with their respective matched filters (i.e. their ``combined" impulse responses) will be automatically identical. For example, the pulses $g_1(t)$ and $g_2(t)$ shown in Fig.~\ref{fig:TBPs} are obtained by convolving the root-raised-cosine pulse~$w(t)$ with two different nonlinear chirps. While $w(t)$, $g_1(t)$, and $g_2(t)$ have significantly different TBPs, their convolutions with the respective matched filters produce the same raised-cosine pulse~$(w\!\ast\! w)(t)$.

We would like to mention in passing that the same approach can be used to construct multidimensional pairs of matched filters with identical spectral characteristics but significantly different time and/or spatial supports. Such filters, for example, can be spatial 2D ($g_i(x,y)$) and/or spatio-temporal 3D ($g_i(x,y,t)$) filters for image and video processing. This is illustrated in Fig.~\ref{fig:TBPs 2D} for 2D filters.

\subsection{Convolution with Large-TBP Filter as Gaussian Mimic Function}
Fig.~\ref{fig:ppileup} illustrates how the pileup effect can be used to obscure (e.g. to mimic as Gaussian or sub-Gaussian) a large-PAPR (super-Gaussian) transmitted signal, while fully recovering its distinct temporal and amplitude structure in the receiver. In this example, convolution of the pulse train with a large-TBP filter in the transmitter ``hides" its original structure, and the pulses with larger TBPs perform this more effectively. This can be seen in Fig.~\ref{fig:ppileup} from both the time-domain traces and the normal probability plots shown in the lower left corner. For a sufficiently large TBP, the distribution of the filtered pulse train becomes effectively Gaussian, making it impossible to distinguish between the individual pulses.

The filters $g_i(t)$ in Fig.~\ref{fig:ppileup} are obtained by filtering the root-raised-cosine pulse $w(t)$ with different all-pass filters, and thus they have the same frequency responses. While $w(t)$ and $g_i(t)$ have significantly different TBPs, their convolutions with the respective matched filters produce the same raised-cosine pulse $(w\!\ast\! w)(t)$. Hence, in the receiver, filtering with a respective matched filter effectively restores the train's original temporal and amplitude structure.

\section{Pulse Trains for Low-SNR Communications} \label{sec:pulse trains for communications}
For sufficiently low pulse rate~${\mathcal{R}}$ (e.g. below half of the bandwidth for~${\rm TBP}\!=\!1$), the PAPR of a pulse train is inversely proportional to~${\mathcal{R}}$, and the magnitude of the pulses in a train of a given power can be made arbitrarily large by reducing the pulse rate. Thus a pulse train consisting of pulses with a small TBP can be effectively used for low-SNR communications, when the Shannon's upper limit on the channel capacity~\cite{Shannon49communication} is itself below the bandwidth.

For the most effective use of the pileup effect for conversion of a high-PAPR pulse train with a distinct, super-Gaussian temporal and amplitude structure into an effectively Gaussian signal, by filtering the train with a large-TBP filter, the pulse train needs to be randomized. This can be accomplished by randomizing the amplitude of the pulses in the train, their arrival times, or both. The ways in which the pulse train is randomized affect the ways in which the information can be encoded and retrieved. For example, if the timing structure of the pulse train is known, synchronous pulse detection can be used. Otherwise, one may need to employ an asynchronous pulse detection (e.g. pulse counting). This, in turn, affects the capacity of the channel.

\begin{figure}[!b]
  \centering
  \includegraphics[width=.666\linewidth]{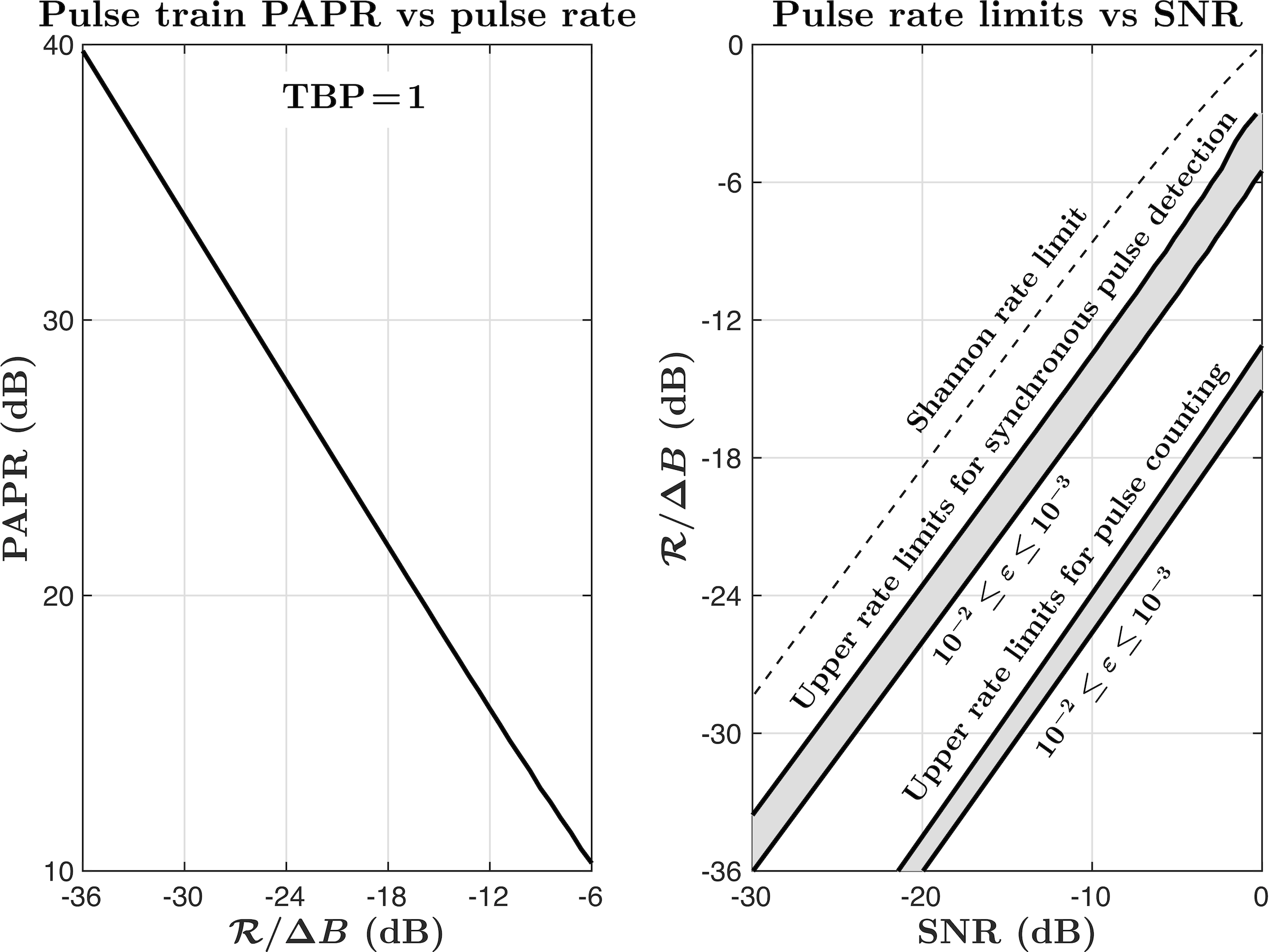}
  \caption{Relations among rate, PAPR, and SNR in pulse train used for low-SNR communications. For ${\rm TBP}\!=\!1$ and $10^{-2}\!\le\!\varepsilon\!\le\!10^{-3}$, ``raw" rate limits for detectible pulses of equal magnitudes vary from few percent (for pulse counting) to about half of Shannon rate (for synchronous pulse detection).}
  \Description{Relations among rate, PAPR, and SNR in pulse train used for low-SNR communications. For ${\rm TBP}\!=\!1$ and $10^{-2}\!\le\!\varepsilon\!\le\!10^{-3}$, ``raw" rate limits for detectible pulses of equal magnitudes vary from few percent (for pulse counting) to about half of Shannon rate (for synchronous pulse detection).}
  \label{fig:pulse rates}
\end{figure}

\subsection{Pulse Counting vs. Synchronous Pulse Detection} \label{subsec:detection vs counting}
Let us consider a pulse train consisting of pulses with the bandwidth~$\Delta{B}$ and a small TBP, so that a single large-magnitude peak in a pulse dominates, and assume that the arrival rate~${\mathcal{R}}$ of the pulses is sufficiently small so that pileup is negligible (e.g. ${\mathcal{R}}\ll \half\Delta{B}/{\rm TBP}$). When the arrival time of a pulse with the peak amplitude~$A>0$ is known, the probability of detecting this pulse as positive in the presence of Gaussian noise with zero mean and the variance~$\sigma_{\rm n}^2$ can be expressed as ${\half\erfc\left(\frac{-A}{\sigma_{\rm n}\sqrt{2}}\right)}$. Then the pulses with the amplitude ${A>\sigma_{\rm n}\sqrt{2}\erfc^{-1}(2\varepsilon)}$ will have a pulse identification error rate smaller than~$\varepsilon$. For example, $\varepsilon\lesssim 1.3\!\times\! 10^{-3}$ for $A\gtrsim 3\sigma_{\rm n}$, and $\varepsilon\lesssim 3.2\!\times\! 10^{-5}$ for $A\gtrsim 4\sigma_{\rm n}$.

In pulse counting, a pulse is detected when it crosses a certain threshold. A {\itshape false positive\/} detection occurs when such crossing is entirely due to noise, and a {\itshape false negative\/} detection happens when a pulse affected by the noise fails to cross the threshold. For a positive threshold~$\alpha_+>0$, the false negative rate will be smaller than~$\varepsilon$ if the amplitude of a pulse is ${A>\alpha_+ + \sigma_{\rm n}\sqrt{2}\erfc^{-1}(2\varepsilon)}$.

As shown in~\cite{Rice44and45mathematical, Nikitin98a}, for a filtered noise with zero mean and the variance~$\sigma_{\rm n}^2$, its rate of up-crossing the threshold~$\alpha_+>0$ can be expressed as ${{\mathcal{R}}_{\rm max}\exp \left( -\half(\alpha_+/\sigma_{\rm n})^2\right)}$, where the {\itshape saturation rate\,}~${\mathcal{R}}_{\rm max}$ is determined entirely by the filter's frequency response. Then, for the average pulse arrival rate~${\mathcal{R}}$, the threshold value needs to be $\alpha_+ > \sigma_{\rm n} \left[ -2\ln(\varepsilon {\mathcal{R}}/{\mathcal{R}}_{\rm max}) \right]^\half$ in order to keep the false positive rate below~$\varepsilon$. For example, for ${\mathcal{R}}/{\mathcal{R}}_{\rm max}=1/10$, $\alpha_+ \gtrsim 4.3\sigma_{\rm n}$ for $\varepsilon=10^{-3}$, and $\alpha_+ \gtrsim 4.8\sigma_{\rm n}$ for $\varepsilon=10^{-4}$. Note that for an ideal ``brick wall" lowpass filter with the bandwidth~$\Delta{B}$ the saturation rate ${\mathcal{R}}_{\rm max}=\Delta{B}/\sqrt{3}$~\cite{Rice44and45mathematical}. Hence, for example, for a root-raised-cosine or a raised-cosine filter ${\mathcal{R}}_{\rm max}\approx (2T_{\rm s}\sqrt{3})^{-1}$, where~$T_{\rm s}$ is the reciprocal of the symbol-rate parameter of the filter.

For a pulse rate~${\mathcal{R}}$ that is sufficiently smaller than ${\mathcal{R}}_0\!=\! \half\Delta{B}/{\rm TBP}$, the PAPR of a train of equal-magnitude pulses is inversely proportional to~${\mathcal{R}}$. This is illustrated in the left panel of Fig.~\ref{fig:pulse rates} for a pulse train consisting of root-raised-cosine pulses. Then, for a given signal-to-noise ratio (SNR) of a pulse train affected by additive Gaussian noise, and for a given error rate constraint~$\varepsilon$, the pulse rate needs to be sufficiently small to ensure the pulse detection with the error rate below~$\varepsilon$. This is illustrated in the right panel of Fig.~\ref{fig:pulse rates}, for both pulse counting and synchronous pulse detection, for $10^{-2}\!\le\!\varepsilon\!\le\!10^{-3}$ and a pulse train consisting of root-raised-cosine pulses. For example, as shown in this panel, for the SNR equal to~$-10\,$dB the upper rate limits for $10^{-2}\!\le\!\varepsilon\!\le\!10^{-3}$ are approximately~$(2.8-4.1)\times\! 10^{-3}\Delta{B}$ for pulse counting, and~$(2.5-4.5)\times\! 10^{-2}\Delta{B}$  for synchronous pulse detection, where~$\Delta{B}$ is the bandwidth of the signal. For comparison, the Shannon upper limit on channel capacity~\cite{Shannon49communication} is shown by the dashed line.

While the rate limit for pulse counting is approximately an order of magnitude lower than for synchronous pulse detection, pulse counting does not rely on any {\itshape a priori\/} knowledge of pulse arrival times, and can be used as a backbone method for pulse detection. Thus it is used in all subsequent examples of this paper. In practice, both pulse counting and synchronous pulse detection can be used in combination. For example, given a constraint on the total power of the pulse train, counting of relatively rare, higher-amplidude pulses can be used to establish the timing patterns for synchronization, and synchronous detection of smaller, more frequent pulses can be used for a higher data rate.

\begin{figure}[!b]
  \centering
  \includegraphics[width=.666\linewidth]{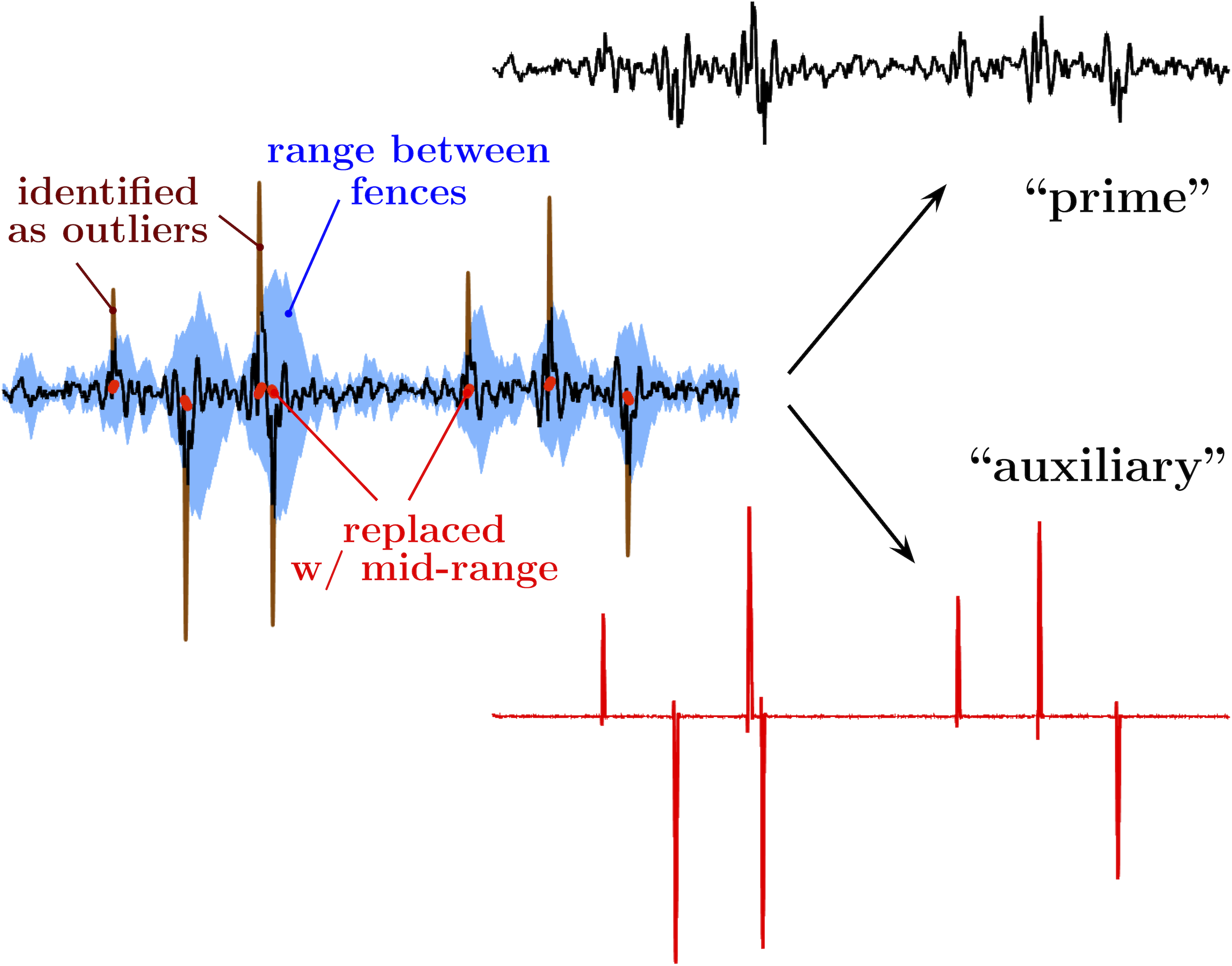}
  \caption{Intermittently Nonlinear Filtering (INF): Outliers are identified as protrusions outside of fenced range, and their values are replaced by those in mid-range. Otherwise, signal is not affected. ``Auxiliary" output is difference between input and ``prime" INF output.}
  \Description{Intermittently Nonlinear Filtering (INF): Outliers are identified as protrusions outside of fenced range, and their values are replaced by those in mid-range. Otherwise, signal is not affected. ``Auxiliary" output is difference between input and ``prime" INF output.}
  \label{fig:INF}
\end{figure}
\begin{figure}[!t]
  \centering
  \includegraphics[width=.666\linewidth]{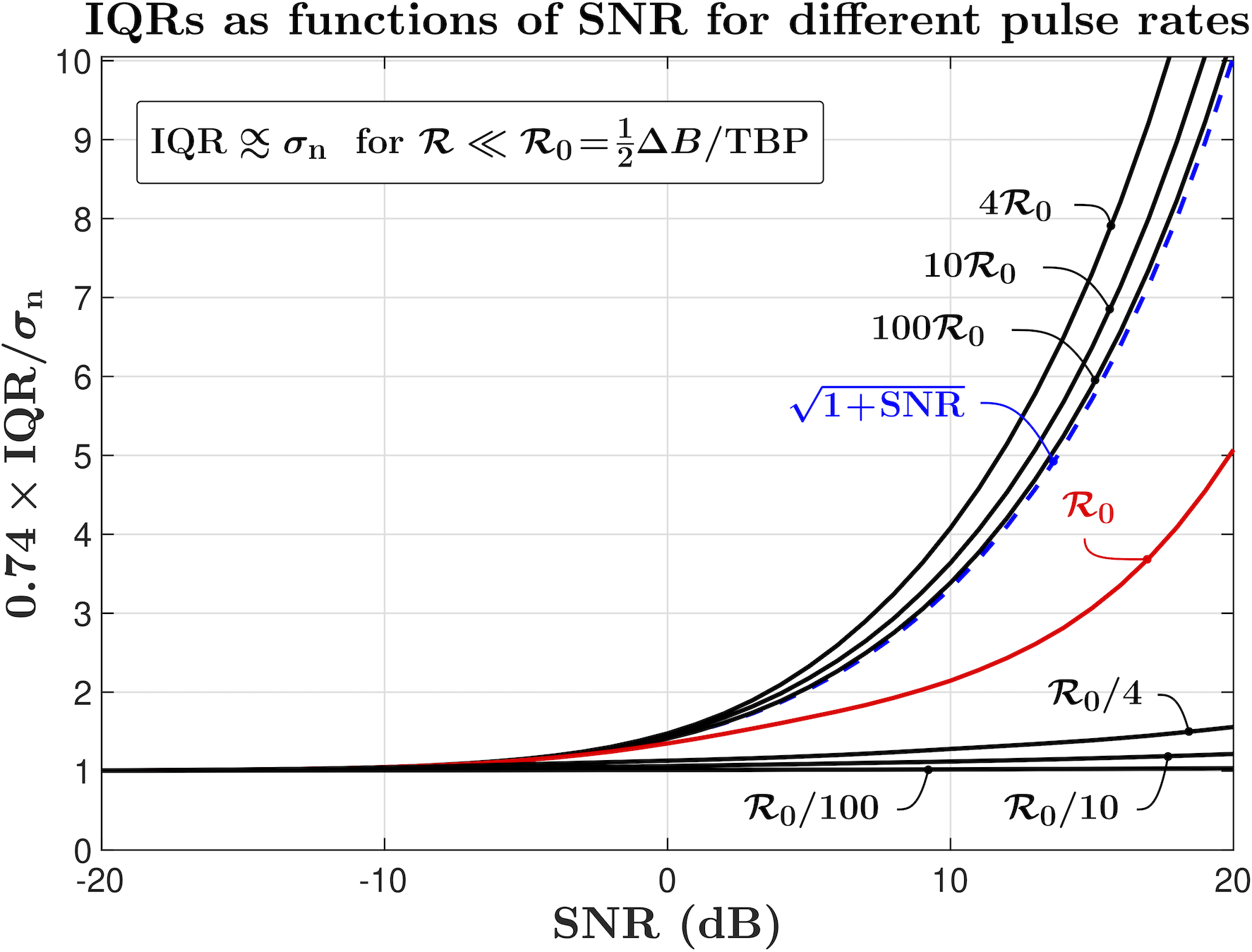}
  \caption{\boldmath For low pulse rates (e.g. ${\mathcal{R}}\!\ll\! \half\Delta{B}/{\rm TBP}$), IQR provides reliable measure of additive Gaussian noise power, $\sigma_{\rm n}\propto {\rm IQR}$. Root-raised-cosine pulses (for which ${\mathcal{R}}_0=(4T_s)^{-1}$) are used in this example. For completeness, IQRs for higher rates are also shown, but details of their change with SNR are not discussed.}
  \Description{For low pulse rates (e.g. ${\mathcal{R}}\!\ll\! \half\Delta{B}/{\rm TBP}$), IQR provides reliable measure of additive Gaussian noise power, $\sigma_{\rm n}\propto {\rm IQR}$. Root-raised-cosine pulses (for which ${\mathcal{R}}_0=(4T_s)^{-1}$) are used in this example. For completeness, IQRs for higher rates are also shown, but details of their change with SNR are not discussed.}
  \label{fig:IQR}
\end{figure}
\begin{figure}[!b]
  \centering
  \includegraphics[width=.75\linewidth]{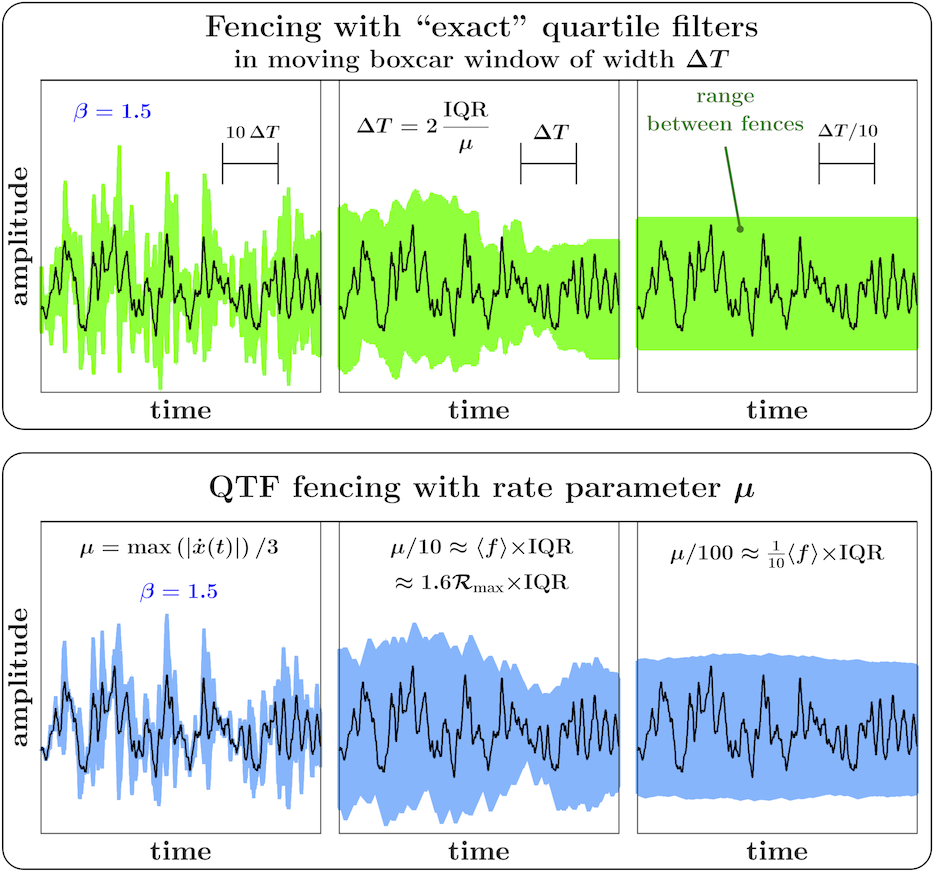}
  \caption{\boldmath Overall behavior of QTF fencing is similar to that with ``exact" quartile filters in moving boxcar window of width~$\Delta{T}=2\times{\rm IQR}/\mu$.}
  \Description{Overall behavior of QTF fencing is similar to that with ``exact" quartile filters in moving boxcar window of width~$\Delta{T}=2\times{\rm IQR}/\mu$.}
  \label{fig:RANKvsQTFs}
\end{figure}

\section{Intermittently Nonlinear Filtering (INF) for Outlier Mitigation and Pulse Counting} \label{sec:INF}
In general, a nonlinear filter is capable of disproportionately affecting spectral densities of signals with distinct temporal and/or amplitude structures even when these signals have the same spectral content. In particular, the separation of a large-PAPR pulse train and a small-PAPR signal can be viewed as either (i)~mitigation of impulsive noise affecting the small-PAPR signal, or (ii)~extraction of impulsive signal from the small-PAPR background. In this paper, a specific type of Intermittently Nonlinear Filters (INF) is used to accomplish either or both tasks. While various INF configurations, their different uses, and the approaches to their analog and/or digital implementations are described elsewhere~\cite{Nikitin19ADiCpatentCIP1, Nikitin19complementary, Nikitin19hidden, Nikitin19quantile, Nikitin18ADiC-ICC}, Fig.~\ref{fig:INF} illustrates their basic concept. In an INF, the upper and the lower fences establish a robust range that excludes high-amplitude pulses while effectively containing the small-PAPR component. The prime INF output simply contains the input signal in which the outliers (i.e. the pulses that protrude from the range) are replaced with mid-range values. This constitutes mitigation of impulsive noise affecting the small-PAPR signal. The auxiliary INF output is the difference between its input and the prime output. This is akin to extraction of impulsive signal from the small-PAPR background (or ``pulse counting").

\subsection{Robust Range/Fencing in INF} \label{subsec:IQR}
For an INF to be effective in separation of small-PAPR and impulsive signals regardless of their relative powers, its range needs to be robust (insensitive) to the pulse train. Favorably, for a mixture of a small-PAPR signal with bandwidth~$\Delta{B}$, and a pulse train with the same bandwidth and the rate sufficiently below ${\mathcal{R}}_0$, when the pileup effect is insignificant, the value of the interquartile range (IQR) of the mixture is insensitive to the power of the pulse train. This is illustrated in Fig.~\ref{fig:IQR} for a pulse train affected by additive Gaussian noise. Thus robust upper ($\alpha_+$) and lower ($\alpha_-$) fences for INF can be constructed as linear combinations of the 1st ($Q_{[1]}$) and the 3rd ($Q_{[3]}$) quartiles of the signal (Tukey's fences~\cite{Tukey77exploratory}) obtained in a moving time window:
\begin{equation} \label{eq:Tukey's range}
  [\alpha_-,\alpha_+] = {\left[Q_{[1]}\!-\!\beta\left(Q_{[3]}\!-\!Q_{[1]}\right)\!,\,Q_{[3]}\!+\!\beta\left(Q_{[3]}\!-\!Q_{[1]}\right)\!\right]},
\end{equation}
where $\alpha_+$, $\alpha_-$, $Q_{[1]}$, and $Q_{[3]}$ are time-varying quantities, and $\beta$ is a scaling parameter of order unity. When an INF is used for pulse counting in the presence of additive Gaussian noise, the particular value of~$\beta$ should be chosen based on the constraint on the relative rate~$\varepsilon$ of false positive detections. Then, as follows from the discussion in Section~\ref{subsec:detection vs counting},
\begin{equation} \label{eq:beta}
\beta \approx 1.05\times \sqrt{ \ln\left( \frac{{\mathcal{R}}_{\rm max}}{\varepsilon {\mathcal{R}}} \right)} \,- \half\,.
\end{equation}
For example, for ${\mathcal{R}}/{\mathcal{R}}_{\rm max}=1/10$, $\beta \approx 2.7$ for $\varepsilon=10^{-3}$, and $\beta \approx 3.1$ for $\varepsilon=10^{-4}$.

\subsection{Quantile Tracking Filters (QTFs) for Robust Fencing} \label{subsec:QTFs}
As a practical matter, Quantile Tracking Filters (QTFs)~\cite{Nikitin19ADiCpatentCIP1, Nikitin19complementary, Nikitin19hidden, Nikitin19quantile, Nikitin18ADiC-ICC} are an appealing choice for such robust fencing in INF, as QTFs are analog filters suitable for wideband real-time processing of continuous-time signals and are easily implemented in analog circuitry. Further, their numerical computations are $\mathcal{O}(1)$ per output value in both time and storage, which also enables their high-rate digital implementations in real time.

In brief, the signal~$Q_q(t)$ that is related to the given input~$x(t)$ by the equation
\begin{equation} \label{eq:QTF eps}
  \frac{\d}{\d{t}}\, Q_q = \mu\, \left[\lim_{\varepsilon\to 0}\Seps(x\!-\!Q_q) + 2q-1\right],
\end{equation}
where $\mu$ is the {\itshape rate parameter\/} and $0\!<\!q\!<\!1$ is the {\itshape quantile parameter\/}, can be used to approximate (``track") the $q$-th~quantile of $x(t)$ for the purpose of establishing a robust range ${[\alpha_-,\alpha_+]}$. In~(\ref{eq:QTF eps}), the {\itshape comparator function\,}~$\Seps(x)$ can be any {\itshape continuous\/} function such that~$\Seps(x)=\sgn(x)$ for~$|x|\gg\varepsilon$, and $\Seps(x)$~changes monotonically from~``$-1$" to~``$1$" so that most of this change occurs over the range~$[-\varepsilon,\varepsilon]$. As discussed in detail in~\cite{Nikitin19quantile}, for a continuous stationary signal~$x(t)$ with a constant mean and a positive IQR, the outputs~$Q_{[1]}(t)$ and~$Q_{[3]}(t)$ of QTFs with a sufficiently small rate parameter~$\mu$ will approximate the~1st and the~3rd quartiles, respectively, of the signal obtained in a moving boxcar time window with the width~$\Delta{T}$ of order~${2\times{\rm IQR}/\mu\gg\langle f\rangle^{-1}}$, where~$\langle f\rangle$ is the average crossing rate of~$x(t)$ with the~1st and the~3rd quartiles of~$x(t)$. Consequently, as illustrated in Fig.~\ref{fig:RANKvsQTFs}, the overall behavior of the QTF fencing for a stationary constant-mean signal with a given IQR would be similar to the fencing with the ``exact" quartile filters in a moving boxcar window $\left[ \theta(t) - \theta(t\!-\!\Delta{T}) \right]/\Delta{T}$, where~${\Delta{T}=2\times{\rm IQR}/\mu}$ and~$\mu$ is the QTF rate parameter. However, for a sampling rate~$F_{\rm s}$, numerical computations of an ``exact" quartile require $\mathcal{O}\left(F_{\rm s}\Delta{T}\log(F_{\rm s}\Delta{T})\right)$ per output value in time, and $\mathcal{O}(F_{\rm s}\Delta{T})$ in storage, becoming prohibitively expensive for high-rate real-time processing.

\begin{figure}[!t]
  \centering
  \includegraphics[width=.5\linewidth]{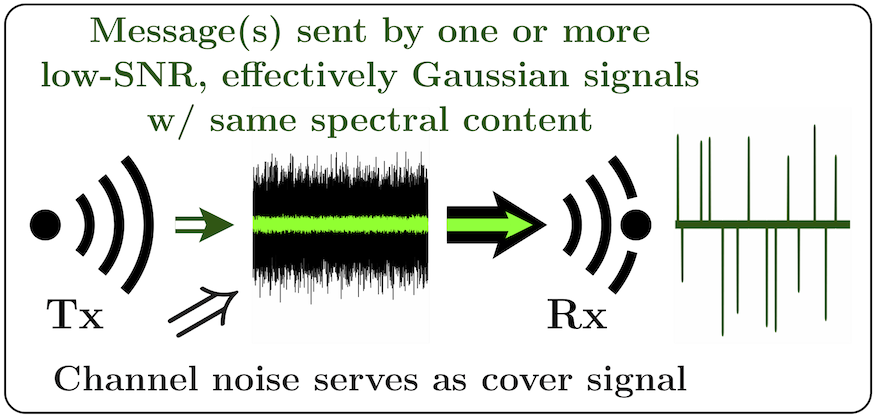}
  \caption{Simplified diagram of first example.}
  \Description{Simplified diagram of first example.}
  \label{fig:TxRx1}
\end{figure}
\begin{figure*}[!b]
  \centering
  \includegraphics[width=\textwidth]{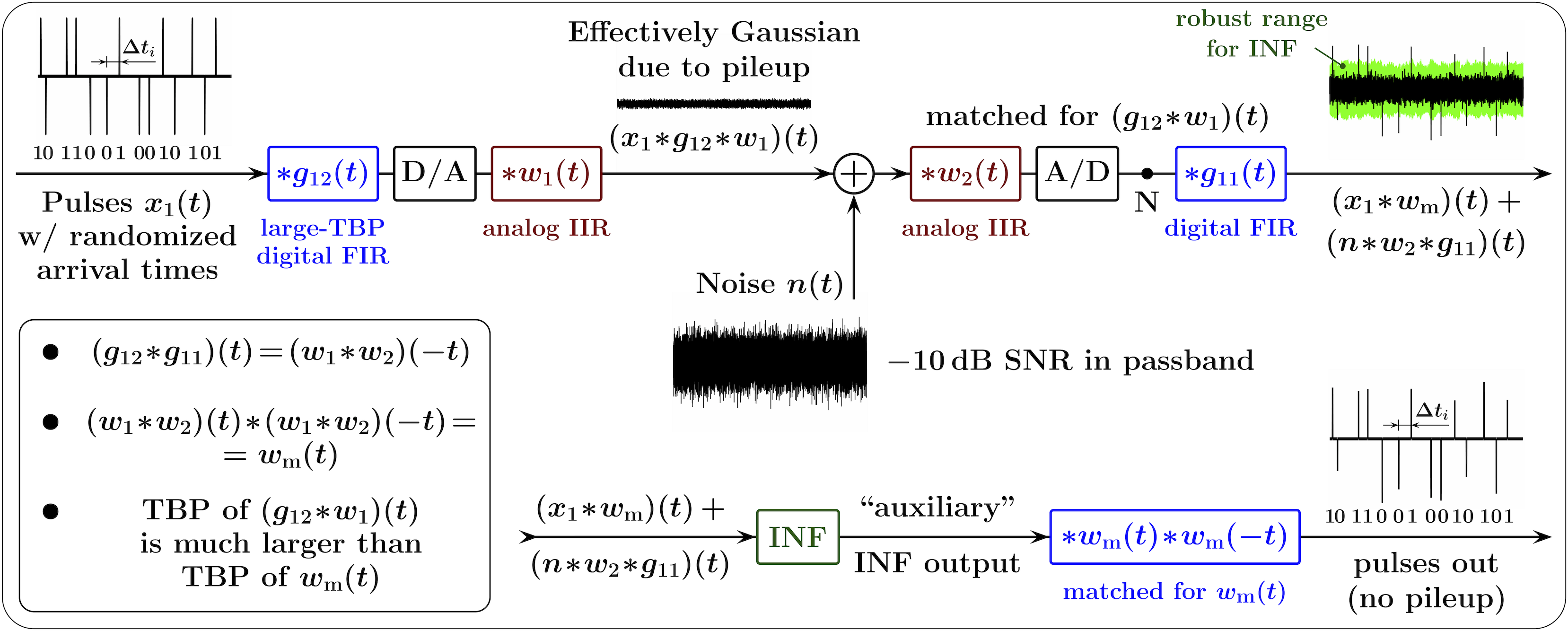}
  \caption{Detailed particular example for basic concept highlighted in Fig.~\ref{fig:TxRx1}.}
  \Description{Detailed particular example for basic concept highlighted in Fig.~\ref{fig:TxRx1}.}
  \label{fig:Imp2Gaussian}
\end{figure*}

\section{Illustrative Examples} \label{sec:examples}
Let us now provide several particular illustrations of utilizing the pileup effect and synergistic combinations of linear and intermittently nonlinear filtering for synthesis of covert and hard-to-intercept communication links.

\subsection{Message Sent by Pulse Train Pretending to be Thermal Noise} \label{subsec:thermal noise}
Fig.~\ref{fig:TxRx1} depicts the basic concept of the first example. The transmitted low-power payload signals are statistically indistinguishable from the Gaussian component of the channel noise (e.g. the thermal noise) observed in the same spectral band, and therefore the channel noise itself serves as a sole cover signal. Further, Fig.~\ref{fig:Imp2Gaussian} provides a detailed particular illustration for the basic concept highlighted in Fig.~\ref{fig:TxRx1}. Here, the message is encoded in a pulse train by both the polarity of the pulses and their interarrival times.\footnote{
We only show the asynchronous detection (pulse counting) performed by obtaining the auxiliary output of an INF. If the rules for the interarrival times are known, synchronous pulse detection can also be used.}
To make this example more realistic, in the transmitter and the receiver we use both digital finite impulse response (FIR) as well as analog (hardware) infinite impulse response (IIR) filters, and include into consideration the respective digital-to-analog (D/A) and analog-to-digital (A/D) conversions. For example, in an underwater acoustic communication system $w_1(t)$ may represent the response of the speaker in the transmitter, and $w_2(t)$ ---  the response of the hydrophone in the receiver.

The channel noise used in the simulation is additive white Gaussian noise (AWGN), and its power is chosen to lead to the $-10\,$dB SNR in the passband of the receiver. Note that the noise can also contain, in addition to Gaussian, a strong outlier component. For example, in underwater acoustic communications it can contain strong impulsive noise produced by snapping shrimp~\cite{Chitre06optimal}. In this case, an additional INF can be deployed before applying the filter~$g_{11}(t)$ in the receiver (e.g. at point~N), to mitigate this noise component and to increase the apparent SNR~\cite{Nikitin19complementary, Nikitin19hidden}.

\begin{figure}[!t]
  \centering
  \includegraphics[width=.5\linewidth]{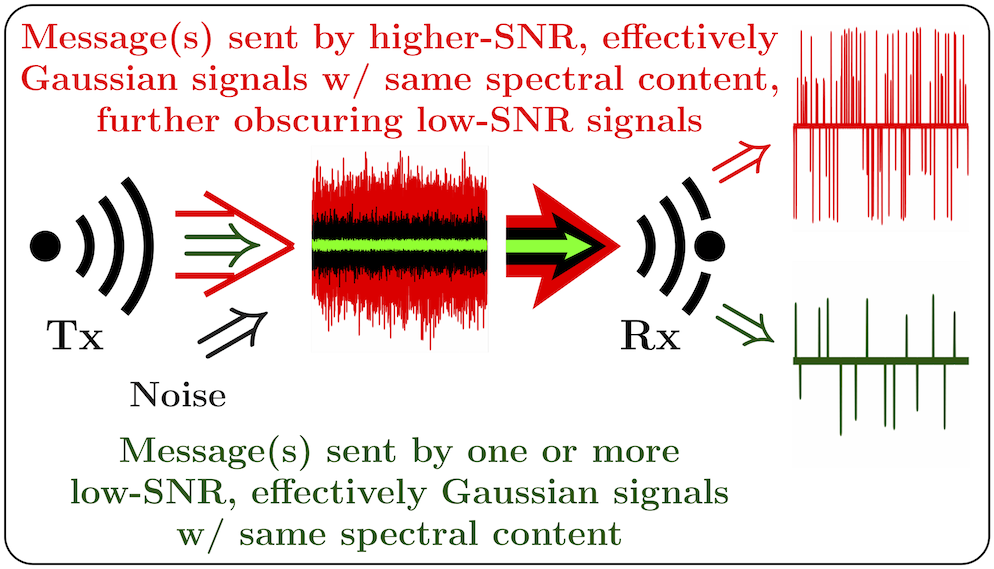}
  \caption{Simplified diagram of second example.}
  \Description{Simplified diagram of second example.}
  \label{fig:TxRx2}
\end{figure}
\begin{figure*}[!t]
  \centering
  \includegraphics[width=\textwidth]{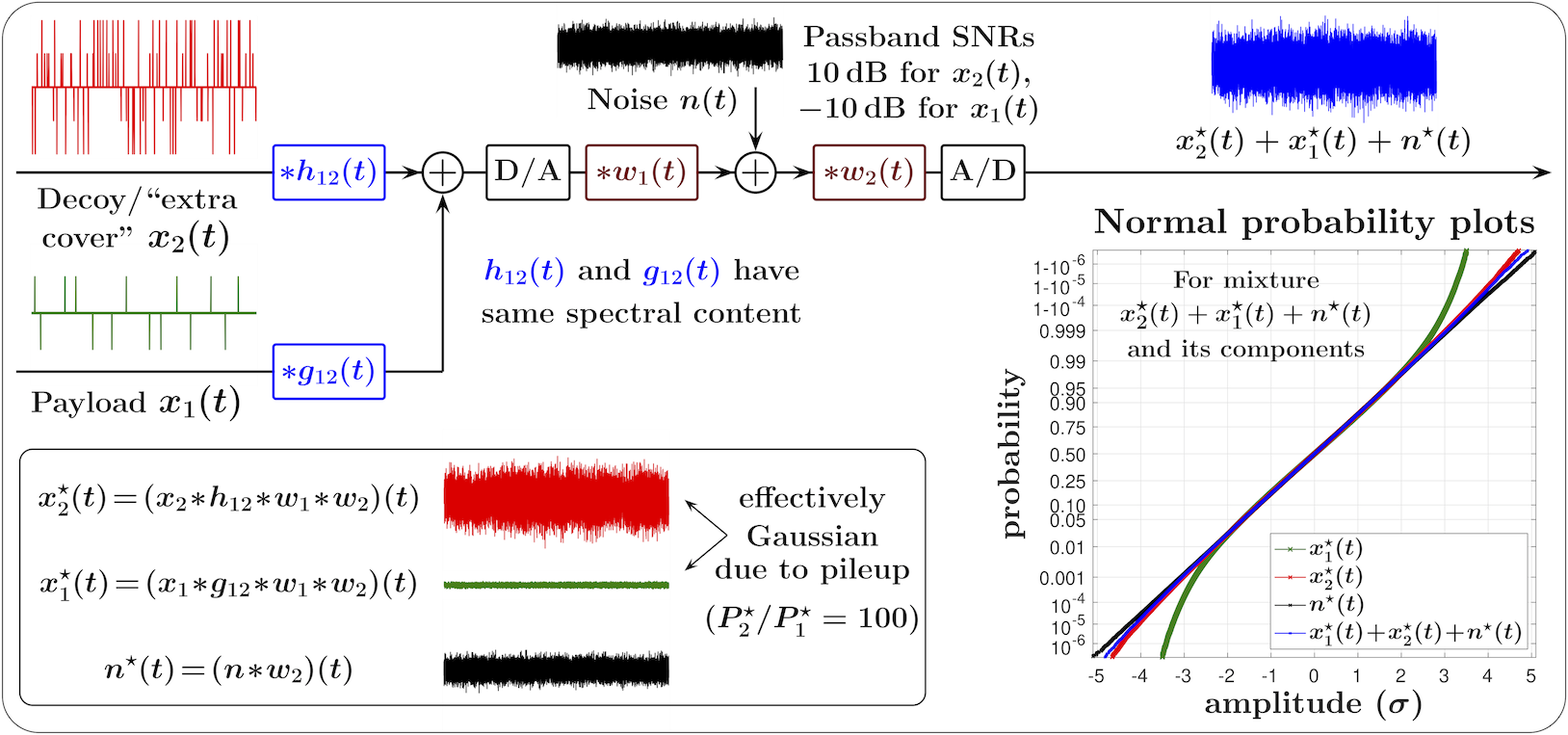}
  \caption{Both high-SNR and low-SNR pulse trains are disguised as Gaussian noises with same spectral content. In this example, time duration of $g_{12}(t)$ is not much larger than average time interval between pulses in $x_1(t)$, and thus $x^\star_1(t)$ is slightly sub-Gaussian.}
  \Description{Both high-SNR and low-SNR pulse trains are disguised as Gaussian noises with same spectral content. In this example, time duration of $g_{12}(t)$ is not much larger than average time interval between pulses in $x_1(t)$, and thus $x^\star_1(t)$ is slightly sub-Gaussian.}
  \label{fig:TxRx2 1}
\end{figure*}
\begin{figure*}[!t]
  \centering
  \includegraphics[width=\textwidth]{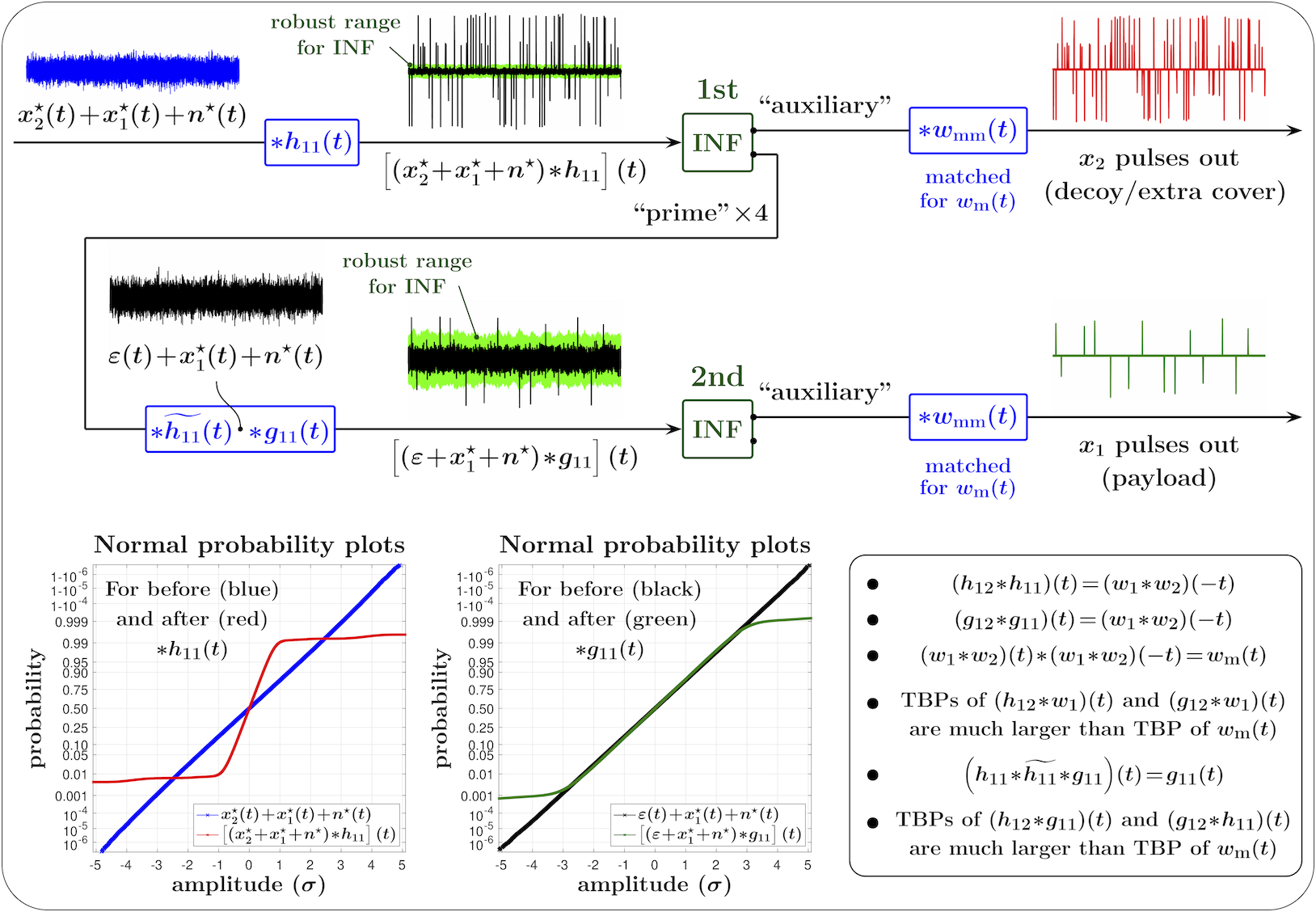}
  \caption{Both high-SNR and low-SNR pulse trains are recovered in receiver. First INF accomplishes both recovery of high-SNR pulse train and its removal from mixture.}
  \Description{Both high-SNR and low-SNR pulse trains are recovered in receiver. First INF accomplishes both recovery of high-SNR pulse train and its removal from mixture.}
  \label{fig:TxRx2 2}
\end{figure*}

\subsection{ Further Obscuring Low-SNR Payloads} \label{subsec:obfuscation}
For a stego pulse train with a given rate, further increasing the power of the channel noise (say, by 10\,dB) can make the pulse train undetectable. For example, when the pulse rate is higher than the Shannon limit for the given SNR, neither synchronous nor asynchronous detection would be possible (see Section~\ref{subsec:detection vs counting}). However, such increase in the channel noise power can be accomplished by an additional pulse train, simply disguised as Gaussian. Then an INF in the receiver, in combination with the respective ``de-mimicking" filter, can effectively remove this additional noise, enabling the detection of the low-power payload. In addition, the higher-power pulse train can itself carry a lower-security (or decoy) message, and/or the timing information that enables synchronous pulse detection in the stego pulse train. Recovering this information from the ``extra cover" signal would still require knowledge of the respective mimic filter used by the transmitter. This concept is schematically illustrated in Fig.~\ref{fig:TxRx2}, and Figs.~\ref{fig:TxRx2 1} and~\ref{fig:TxRx2 2} provide its detailed walk-through example. Note that even after the effective removal of the higher-SNR pulse train from the mixture (by the first INF), the stego message is still Gaussian, and still hidden behind the channel noise (and the remainder of the decoy/timing/``extra cover" signal). Thus its recovery still requires knowledge of the second mimic filter ($g_{12}(t)$) used by the transmitter.

\begin{figure}[!t]
  \centering
  \includegraphics[width=.666\linewidth]{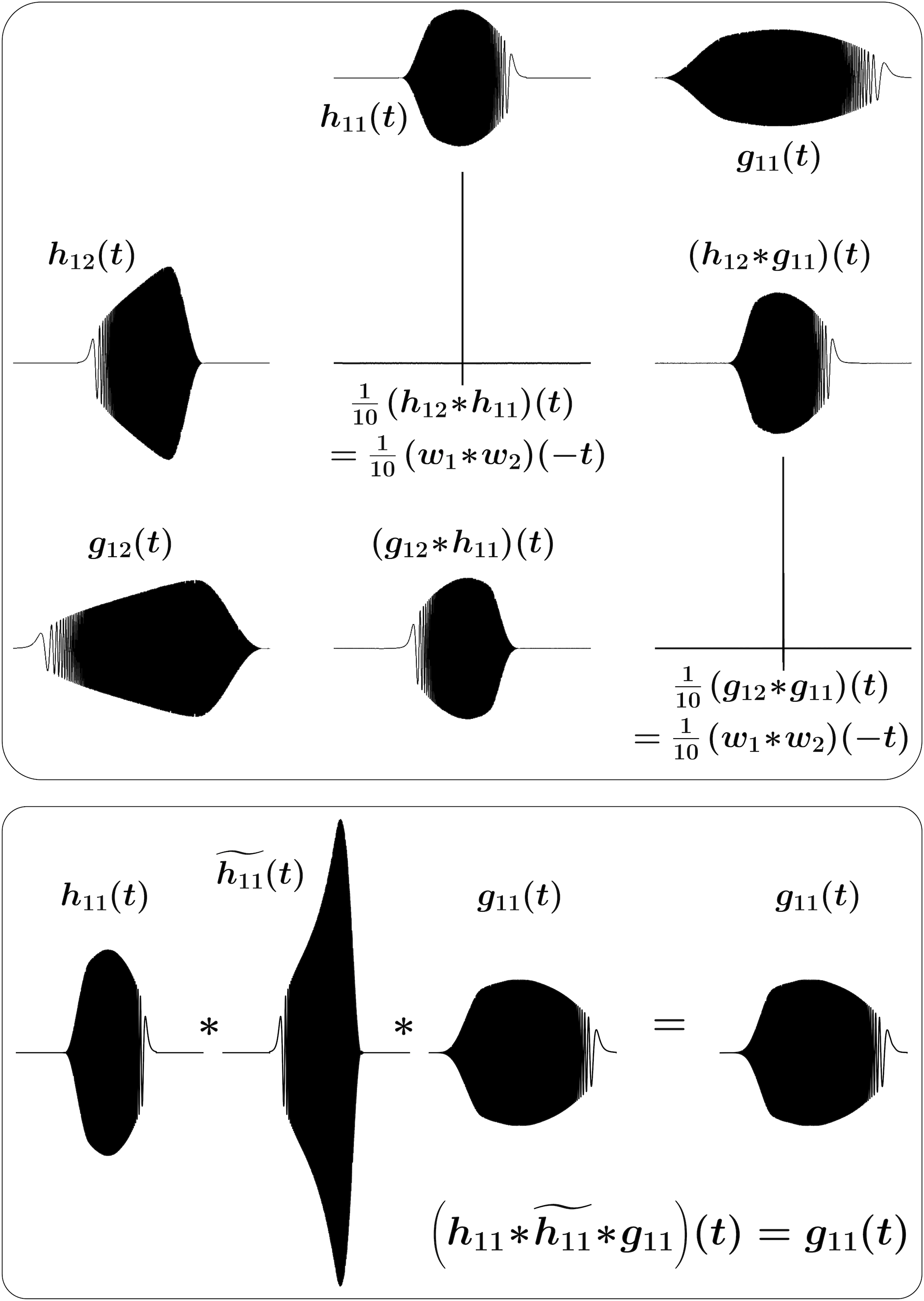}
  \caption{Impulse responses of filters used in Figs.~\ref{fig:TxRx2 1} and~\ref{fig:TxRx2 2}, and their convolutions.}
  \Description{Impulse responses of filters used in Figs.~\ref{fig:TxRx2 1} and~\ref{fig:TxRx2 2}, and their convolutions.}
  \label{fig:TxRx2 filters}
\end{figure}

\subsubsection{Filter properties} \label{subsubsec:filter pairs}
The main properties of the filters used in this example are listed in the lower right panel of Fig.~\ref{fig:TxRx2 2}, and the impulse responses of these filters and their convolutions are illustrated in Fig.~\ref{fig:TxRx2 filters}.

In construction of these filters, we used the approach briefly outlined in Section~\ref{subsec:TBP in context}. In general, given the smallest-TBP filter~$g_0(t)$ with a particular frequency response, one can construct a great variety of filters~$g_i(t)$ with the same frequency response but much larger TBPs (e.g., orders of magnitude larger). These filters can be constructed in such a way that (i)~their combined matched responses are equal to each other, $g_i(t)\ast g_i(-t)=g_j(t)\ast g_j(-t)$ for any $i$ and $j$, and have a small TPB, but (ii)~the convolutions of any $g_i(t)$ with itself (for $i\ne 0$), or with $g_j(\pm t)$ (for~$i\ne j$) have large TBPs. For a given ``seed" pulse~$g_0(t)$, perhaps the easiest way to construct a pulse~$g_i(t)$ with a different TBP is to filter~$g_0(t)$ with an all-pass filter, for example, a linear or nonlinear chirp with a flat frequency response.

\begin{figure}[!b]
  \centering
  \includegraphics[width=.5\linewidth]{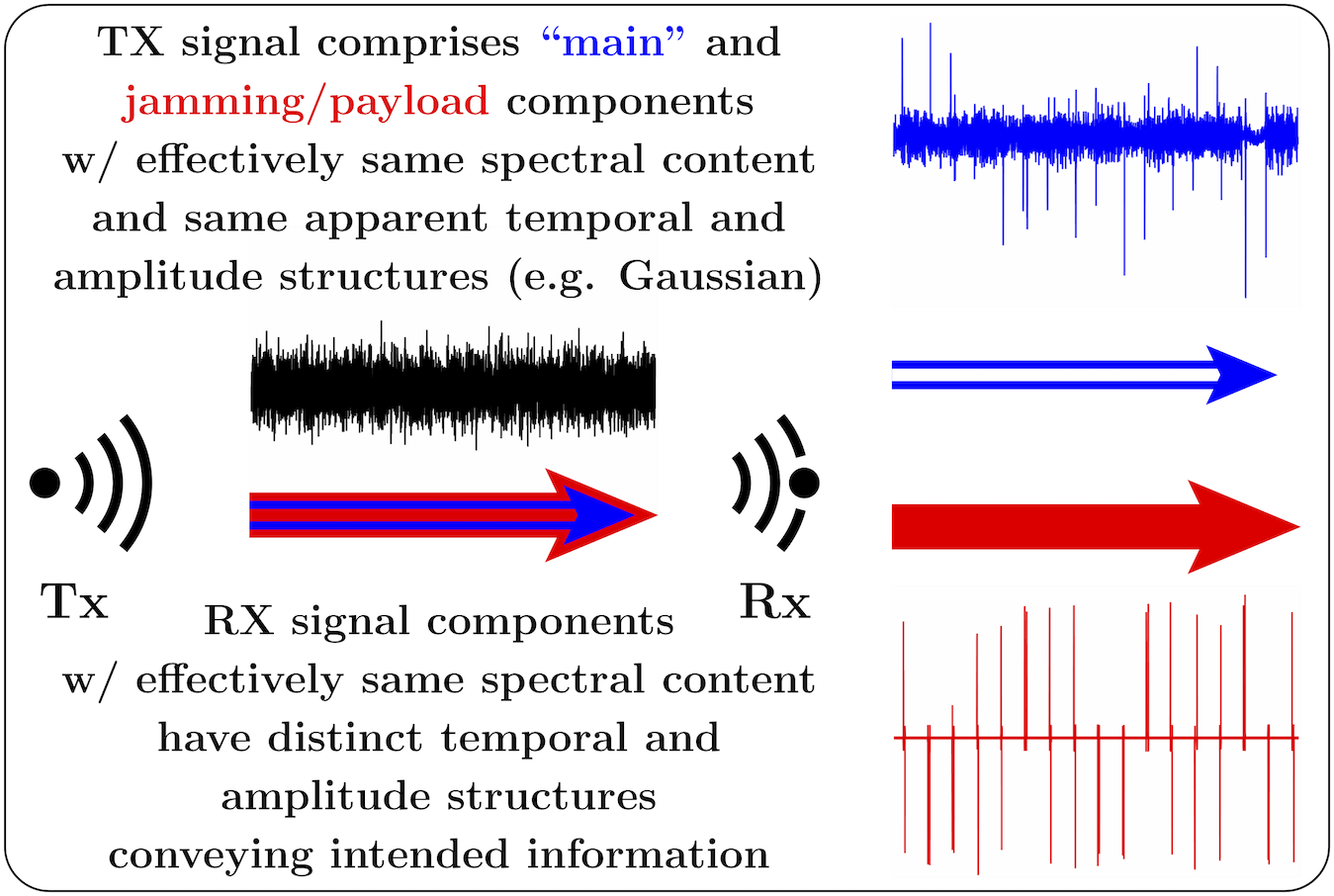}
  \caption{Basic concept of ``friendly in-band jamming."}
  \Description{Basic concept of ``friendly in-band jamming."}
  \label{fig:friendly jamming concept}
\end{figure}
\begin{figure}[!t]
  \centering
  \includegraphics[width=.75\linewidth]{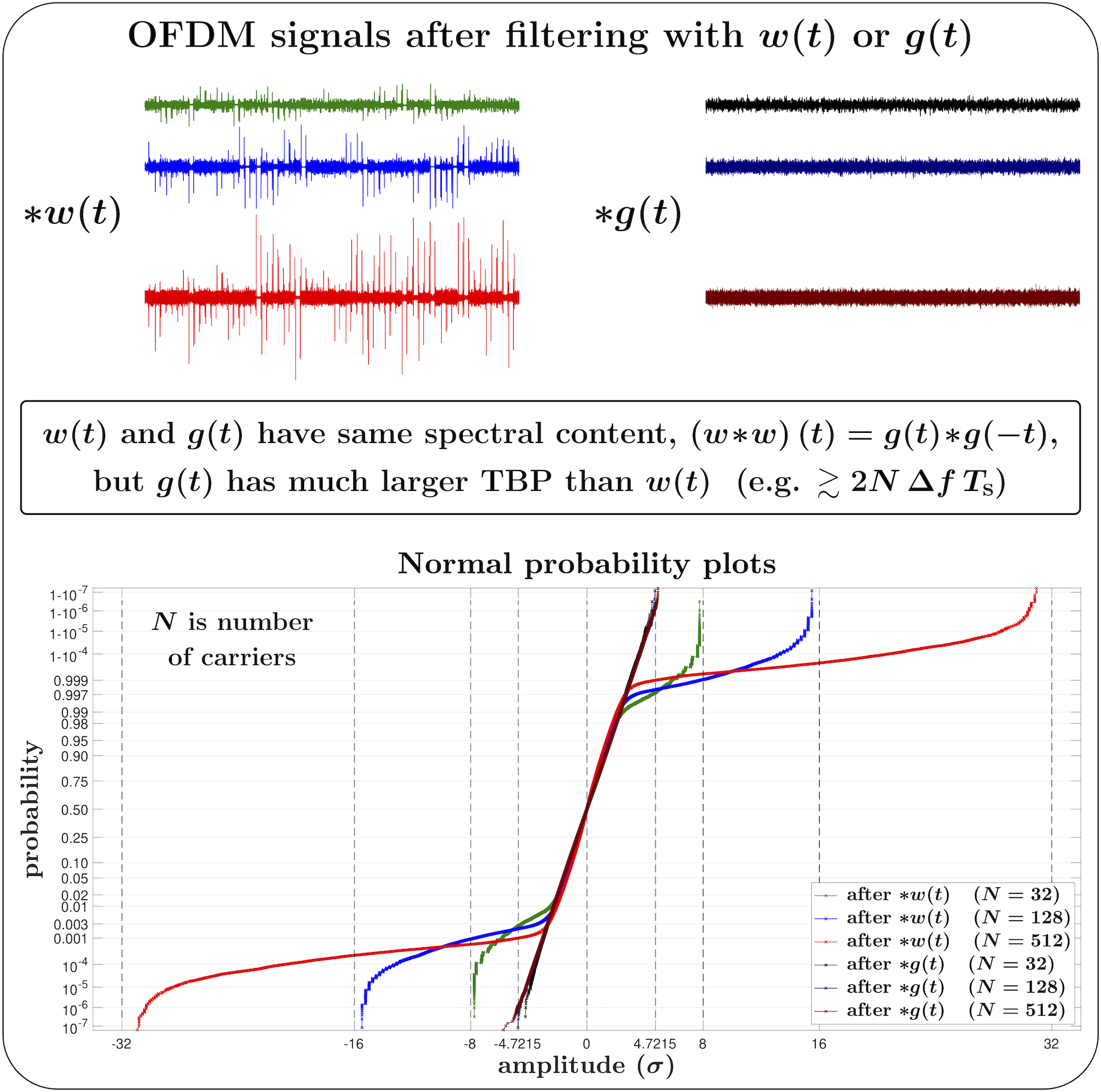}
  \caption{OFDM PAPR reduction by large-TBP filter.}
  \Description{OFDM PAPR reduction by large-TBP filter.}
  \label{fig:OFDM PAPR}
\end{figure}
\begin{figure*}[!t]
  \centering
  \includegraphics[width=\textwidth]{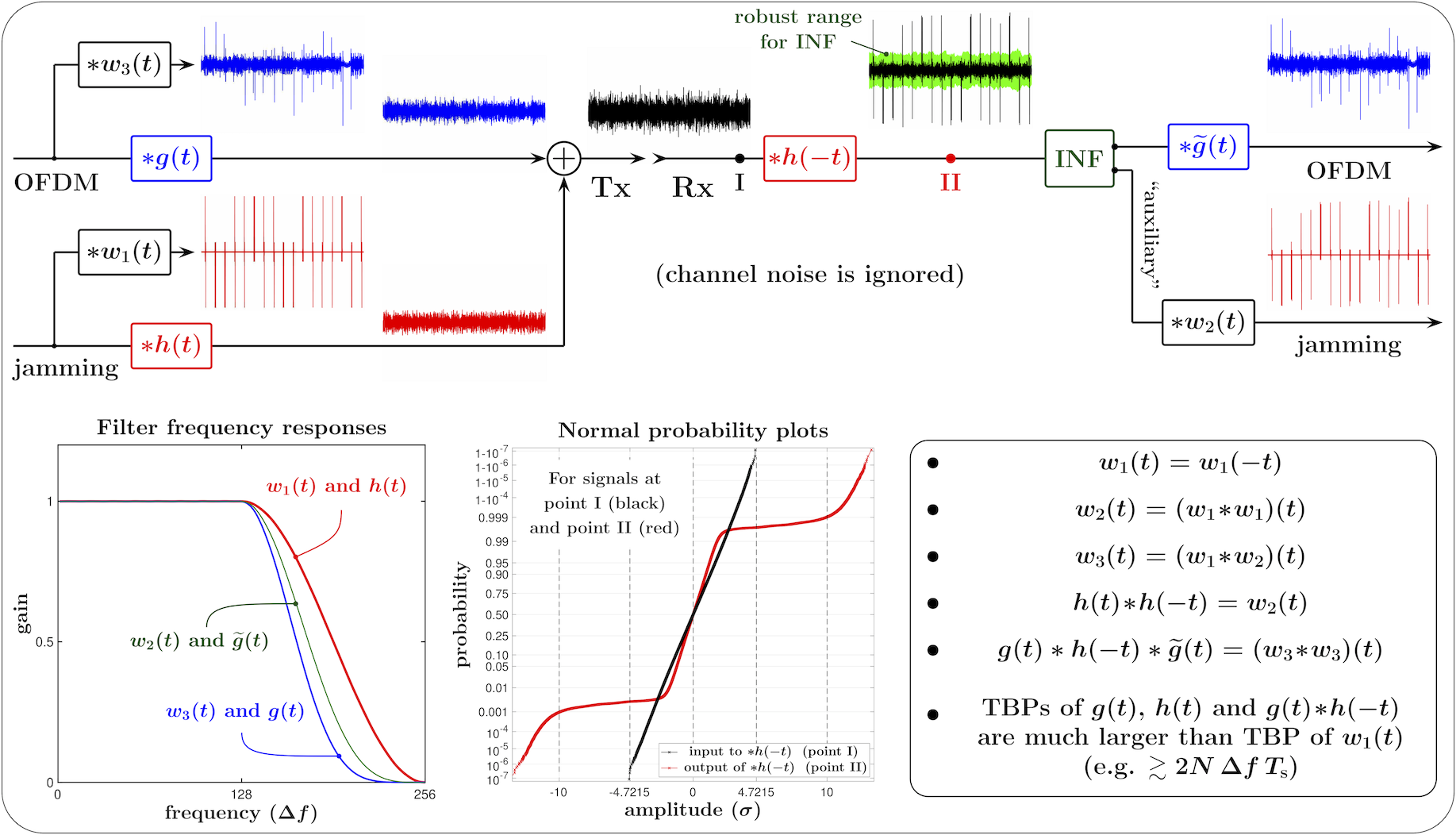}
  \caption{Friendly in-band jamming of OFDM signal. Combination of linear and nonlinear filtering in receiver is used for effective separation of OFDM and ``friendly jamming" signals, although both signals in received mixture have effectively same spectral characteristics and temporal and amplitude structures, and there are no explicit differences in their temporal allocations.}
  \Description{Friendly in-band jamming of OFDM signal. Combination of linear and nonlinear filtering in receiver is used for effective separation of OFDM and ``friendly jamming" signals, although both signals in received mixture have effectively same spectral characteristics and temporal and amplitude structures, and there are no explicit differences in their temporal allocations.}
  \label{fig:OFDM jamming}
\end{figure*}

\subsection{Friendly In-Band Jamming} \label{subsec:friendly jamming}
In our third example, the main message is transmitted using one of the existing communication protocols, but its temporal and amplitude structure is obscured by employing a large-TBP filter in the transmitter, e.g., made to be effectively Gaussian. This alone provides a certain level of security, since the intersymbol interference becomes excessively large and the signal cannot be recovered in the receiver without the knowledge of the mimic filter. In addition, a jamming pulse train, disguised as Gaussian by another (and different) large-TBP filter, is added to the main signal. This jamming signal has effectively the same spectral content as the main signal, and its power is sufficiently large (e.g. similar to the main signal) so that the main signal is unrecoverable even if the first mimic filter is known. In the receiver, the jamming pulse train is removed from the mixture (and recovered, if it itself contains information), enabling the subsequent recovery of the main message. This concept is schematically illustrated in Fig.~\ref{fig:friendly jamming concept}.

\subsubsection{OFDM PAPR Reduction} \label{subsubsec:PAPR}
In addition to improved security, applying a large-TBP filter to the main signal reduces PAPR of large-crest-factor signals such as those in orthogonal frequency-division multiplexing (OFDM), as illustrated in Fig.~\ref{fig:OFDM PAPR}. Here, the simulated OFDM signals are generated without restrictions of the proportion of ``ones" and ``zeros" in a symbol, and thus they have the maximum achievable PAPRs (i.e. $2N$, where~$N$ is the number of carriers).

\subsubsection{Walk-Through Example} \label{subsubsec:walk-through example}
In Fig.~\ref{fig:OFDM jamming}, the main signal is a high-PAPR OFDM signal, and the jamming signal is a high-PAPR impulse train with the spectral content in an effectively the same band (see the frequency responses of the filters in the lower left panel of Fig.~\ref{fig:OFDM jamming}). After the filtering with large-TBP filters $g(t)$ and $h(t)$, respectively, both the OFDM and the jamming signals become effectively Gaussian, and so does their mixture that is being transmitted and received (see the black line in the normal probability plots shown in the lower middle panel of Fig.~\ref{fig:OFDM jamming}).\footnote{In this example, the channel noise is assumed to be relatively small and is not shown.} However, applying a filter matched for~$h(t)$ in the receiver restores the high-PAPR structure of the jamming signal (see the red line in the normal probability plots), while the OFDM component remains Gaussian. Subsequently, the INF accomplishes both the mitigation of the jamming pulse train affecting the OFDM component and the extraction of the jamming signal. Applying the filter~$\widetilde{g}(t)$ to the prime INF output effectively restores the original high-PAPR OFDM signal. If desired, the jamming pulse train is restored by applying the filter~$w_2(t)$ to the auxiliary INF output. 

The main properties of the filters used in this example are listed in the lower right panel of Fig.~\ref{fig:OFDM jamming}, and their frequency responses are shown in the lower left panel of the figure.

\section{Summary} \label{sec:summary}
The main results of this paper can be summarized as follows:
\begin{enumerate}[wide,nosep]
\item Pileup effect can be used for modifying the temporal and amplitude structure of various non-Gaussian signals, and, in many cases, for making them appear as effectively Gaussian. For example, a highly super-Gaussian pulse train consisting of pulses with random amplitudes and/or interarrival times can be converted into an effectively Gaussian or sub-Gaussian by a convolution with a filter having a sufficiently large time-bandwidth product (TBP). Such ``mimicking" of a pulse train as Gaussian noise can be achieved without modifying the spectral content of the train.
\item Given the smallest-TBP filter~$g_0(t)$ with a particular frequency response, one can construct a great variety of filters~$g_i(t)$ with the same frequency response but much larger TBPs (e.g., orders of magnitude larger). These filters can be constructed in such a way that (i)~their combined matched responses are equal to each other, $g_i(t)\!\ast\! g_i(-t)=g_j(t)\!\ast\! g_j(-t)$ for any $i$ and $j$, and have a small TPB, but (ii)~the convolutions any of $g_i(t)$ with itself (for $i\ne 0$), or with $g_j(\pm t)$ (for~$i\ne j$) have large TBPs. For a given ``seed" pulse~$g_0(t)$, perhaps the easiest way to construct a pulse~$g_i(t)$ with a different TBP is to filter~$g_0(t)$ with an all-pass filter, for example, a linear or nonlinear chirp with a flat frequency response.
\item Matched filter pairs with similar properties (i.e. identical spectral characteristics but significantly different time and/or spatial supports) can also be constructed for multidimensional filters, for example spatial 2D ($g_i(x,y)$) and/or spatio-temporal 3D ($g_i(x,y,t)$) filters for image and/or video processing.
\item For sufficiently low pulse rate~${\mathcal{R}}$ (e.g. below half of the bandwidth for~${\rm TBP}\!=\!1$), the PAPR of a pulse train is inversely proportional to~${\mathcal{R}}$, and the magnitude of the pulses in a train of a given power can be made arbitrarily large by reducing the pulse rate. Thus a pulse train consisting of pulses with a small TBP can be effectively used for low-SNR communications, when the Shannon's upper limit on the channel capacity is itself below the bandwidth. For example, if the timing structure of the pulse train is known, synchronous pulse detection can be used. Then, in the presence of additive Gaussian noise and for a train consisting of equal-magnitude pulses with unit TBP, the pulses with the arrival rates in the 25\% to 50\% range of the Shannon's limit for a given SNR can be detected with the raw error rate in the range $10^{-2}\!\le\!\varepsilon\!\le\!10^{-3}$. Using proper modulation of the pulse train (e.g. in terms of the pulse amplitudes and their interarrival times), and error correction coding, the data rate capacity of a pulse train can be brought closer to the Shannon's limit.
\item When the pulse arrival times are unknown (e.g. the interarrival times are random), the asynchronous pulse detection (pulse counting) can be used. In pulse counting, a pulse is detected when it crosses a certain threshold, and this threshold needs to be sufficiently high to ensure a low rate of false positive detections. Therefore, to ensure a comparable to the synchronous pulse detection error rate, for pulse counting the pulse arrival rate needs to be reduced by about an order of magnitude, down to a few percent of the respective Shannon's rate. For example, to 56--82\,kHz for a 20\,MHz channel at $-10$\,dB SNR and $10^{-2}\!\le\!\varepsilon\!\le\!10^{-3}$, as compared to 500--900\,kHz at the same SNR for synchronous detection. In practice, both pulse counting and synchronous pulse detection can be used in combination. For example, given a constraint on the total power of the pulse train, counting of relatively rare, higher-amplidude pulses can be used to establish the timing patterns for synchronization, and synchronous detection of smaller, more frequent pulses can be used for a higher data rate.
\item When each of two or more (say, $N$) pulse trains consists of identically shaped pulses, then, in general, their mixture cannot be effectively separated back into the individual pulse trains. (That is, unless interference among the trains is negligible and a sufficient information about the pulse arrival times in the individual pulse trains is available.) However, before the mixing, one can filter each of the individual pulse trains with ``its own" large-TBP $g_i(t)$, $i\!=\!1,\ldots,N$, so that the mixture becomes an effectively Gaussian signal due to pileup effect. One can then apply to the mixture the filter $g_i(-t)$ such that the pulse $g_i(t)\!\ast\! g_i(-t)$ has the smallest TBP for the given spectral content, but the convolutions $g_j(t)\!\ast\! g_i(-t)$ for $j\ne i$ will still have sufficiently large TBPs so that the mixture of the remaining $N\!-\!1$ pulse trains remains a Gaussian signal. This filtered mixture can then be viewed as (i)~a large-PAPR pulse train affected by additive Gaussian noise, or as (ii)~an effectively Gaussian signal affected by impulsive noise.
\item In general, a nonlinear filter is capable of disproportionately affecting spectral densities of signals with distinct temporal and/or amplitude structures even when these signals have the same spectral content. In particular, the separation of a large-PAPR pulse train and a small-PAPR signal can be viewed as either (i)~mitigation of impulsive noise affecting the small-PAPR signal, or (ii)~extraction of impulsive signal from the small-PAPR background. In this paper, a specific type of Intermittently Nonlinear Filters (INF) is used to accomplish either or both tasks. In such filtering, the upper and the lower fences establish a robust range that excludes high-amplitude pulses while effectively containing the small-PAPR component. The prime output of an INF simply contains the input signal in which the outliers (i.e. the pulses that protrude from the range) are replaced with mid-range values. This constitutes mitigation of impulsive noise affecting the small-PAPR signal. The auxiliary INF output is the difference between its input and the prime output. This is akin to extraction of impulsive signal from the small-PAPR background (or ``pulse counting").
\item For an INF to be effective in separation of small-PAPR and impulsive signals regardless of their relative powers, its range needs to be robust (insensitive) to the pulse train. Favorably, for a mixture of a small-PAPR signal with bandwidth~$\Delta{B}$, and a pulse train with the same bandwidth and the rate sufficiently below ${\mathcal{R}}_0\!=\! \half\Delta{B}/{\rm TBP}$, when the pileup effect is insignificant, the value of the interquartile range (IQR) of the mixture is insensitive to the power of the pulse train. Thus robust upper and lower fences for INF can be constructed as linear combinations of the 1st and the 3rd quartiles of the signal (Tukey's fences) obtained in a moving time window. As a practical matter, Quantile Tracking Filters (QTFs) are an appealing choice for such robust fencing in INF, as QTFs are analog filters suitable for wideband real-time processing of continuous-time signals and are easily implemented in analog circuitry. Further, their numerical computations are $\mathcal{O}(1)$ per output value in both time and storage, which also enables their high-rate digital implementations in real time.
\item The very existence of a detectable carrier (cover signal) may be a dead giveaway for the stego payload. For example, a simple presence of a sheet of paper implies the possibility of a message written in invisible ink. Therefore, the best steganography should be ``carrier-less," when the payload is covertly embedded into something ``ever-present." In the physical layer, such ``ideal" and unidentifiable cover signal is the channel noise. Such noise always includes the ever-present thermal noise as one of its components, and may also comprise other (in general, non-Gaussian) natural and/or technogenic (man-made) components. Then, if the stego payload ``pretends" to be Gaussian, and its power is small enough to be well within the natural variations of the channel noise, any physically available band can be used to carry a virtually undetectable covert message.
\item Further, the paper provides several detailed examples of applying the above concepts to synthesis of covert and hard-to-intercept communication links. These examples include (i)~using the channel noise as a sole cover signal for a low-power payload, (ii)~additional obfuscation of a low-power messages by strong decoy and/or auxiliary/timing signals, and (iii)~``friendly" jamming by a signal with the same spectral content as the main signal that uses a standard protocol. All these examples rely on pileup effect for PAPR control, and on combinations of INF and linear filtering for effective separation of statistically indistinguishable, same-spectral-band cover and payload signals.
\item Note that when the channel noise itself contains an outlier component, an INF deployed early in the receiver chain can mitigate such outlier noise, increasing the overall SNR and the throughput capacity of all channels in the receiver. 
\end{enumerate}

In a broader context, the approach outlined in this paper allows for many practical variations, ranging from simple and easily implementable to more elaborate, highly secure multi-level configurations that would require addressing additional implementational challenges.

\begin{acks}
The authors would like to thank
Kendall Castor-Perry (aka The Filter Wizard);
James~E. Gilley of BK Technologies, West Melbourne, FL;
Arlie Stonestreet\,\,II of Ultra Electronics ICE, Manhattan, KS, and
Kyle~D. Tidball of Textron Aviation, Wichita, KS,
for their valuable suggestions and critical comments.
This work was supported in part by Pizzi Inc., Denton, TX 76205 USA.
\end{acks}

\end{document}